\documentclass[a4paper,11pt]{article}
\pdfoutput=1 

\usepackage{jinstpub} 

\usepackage{subcaption}
\usepackage[english]{babel}
\usepackage{xspace}
\newcommand{\GEANTfour} {{\textsc{Geant4}}\xspace}

\title{Performance of a full scale prototype detector at the BR2 reactor for the SoLid experiment}

\collaboration{The SoLid Collaboration}
\author[a]{Y.~Abreu}
\author[i]{Y.~Amhis}
\author[b]{L.~Arnold}
\author[d]{G.~Ban}
\author[a]{W.~Beaumont}
\author[i]{M.~Bongrand}
\author[i]{D.~Boursette}
\author[j]{B.~C.~Castle}
\author[b]{K.~Clark}
\author[k]{B.~Coup\'e}
\author[b]{D.~Cussans}
\author[a,e]{A.~De Roeck}
\author[c]{J.~D'Hondt}
\author[d]{D.~Durand}
\author[h]{M.~Fallot}
\author[k]{L.~Ghys}
\author[h]{L.~Giot}
\author[d]{B.~Guillon}
\author[g]{S.~Ihantola}
\author[a]{X.~Janssen}
\author[k]{S.~Kalcheva}
\author[c]{L.~N.~Kalousis}
\author[k]{E.~Koonen}
\author[f]{M.~Labare}
\author[d]{G.~Lehaut}
\author[i]{L.~Manzanillas}
\author[k]{J.~Mermans}
\author[f]{I.~Michiels}
\author[f,k]{C.~Moortgat}
\author[b,m]{D.~Newbold}
\author[l]{J.~Park}
\author[d]{V.~Pestel}
\author[b]{K.~Petridis}
\author[a]{I.~Pi\~nera}
\author[b]{G.~Pommery}
\author[k]{L.~Popescu}
\author[h]{G.~Pronost}
\author[b]{J.~Rademacker}
\author[f]{D.~Ryckbosch}
\author[j]{N.~Ryder}
\author[g]{D.~Saunders}
\author[i]{M.-H.~Schune}
\author[i,n]{L.~Simard}
\author[g]{A.~Vacheret }
\author[k]{S.~Van Dyck}
\author[c]{P.~Van Mulders}
\author[a]{N.~van Remortel}
\author[a,c]{S.~Vercaemer} 
\author[a]{M.~Verstraeten} 
\author[j,m]{A.~Weber}
\author[h]{F.~Yermia}

\affiliation[a]{Universiteit Antwerpen, Antwerpen, Belgium}
\affiliation[b]{University of Bristol, Bristol, UK}
\affiliation[c]{Vrije Universiteit Brussel, Brussel, Belgium}
\affiliation[d]{Normandie Univ, ENSICAEN, UNICAEN, CNRS/IN2P3, LPC Caen, 14000 Caen, France}
\affiliation[e]{CERN, 1211 Geneva 23, Switzerland}
\affiliation[f]{Universiteit Gent, Gent, Belgium}
\affiliation[g]{Imperial College London, Department of Physics, London, United Kingdom}
\affiliation[h]{SUBATECH, CNRS/IN2P3, Universit\'e de Nantes, Ecole des Mines de Nantes, Nantes, France}
\affiliation[i]{LAL, Univ Paris-Sud, CNRS/IN2P3, Universit\'e Paris-Saclay, Orsay, France}
\affiliation[j]{University of Oxford, Oxford, UK}
\affiliation[k]{SCK-CEN, Belgian Nuclear Research Centre, Mol, Belgium}
\affiliation[l]{Center for Neutrino Physics, Virginia Tech, Blacksburg, Virginia, 24061, USA}
\affiliation[m]{STFC, Rutherford Appleton Laboratory, Harwell Oxford, and Daresbury Laboratory, Warrington, United Kingdom}
\affiliation[n]{Institut Universitaire de France, F-75005 Paris, France}

\emailAdd{nick.vanremortel@uantwerpen.be}

\abstract{
The SoLid collaboration has developed a new detector technology to detect electron anti-neutrinos at close proximity to the Belgian BR2 reactor at surface level. A 288$\,$kg prototype detector was deployed in 2015 and collected data during the operational period of the reactor and during reactor shut-down. Dedicated calibration campaigns were also performed with gamma and neutron sources.

This paper describes the construction of the prototype detector with a high control on its proton content and the stability of its operation over a period of several months after deployment at the BR2 reactor site. All detector cells provide sufficient light yields to achieve a target energy resolution of better than 20\%/$\sqrt{E(\mathrm{MeV})}$. The capability of the detector to track muons is exploited to equalize the light response of a large number of channels to a precision of 3\% and to demonstrate the stability of the energy scale over time. Particle identification based on pulse-shape discrimination is demonstrated with calibration sources.
Despite a lower neutron detection efficiency due to triggering constraints, the main backgrounds at the reactor site were determined and taken into account in the shielding strategy for the main experiment. The results obtained with this prototype proved essential in the design optimization of the final detector.

\vspace{1cm}
{\em 
This paper is dedicated to our SCK\raisebox{-0.8ex}{\scalebox{2.8}{$\cdot$}}CEN colleague, Edgar Koonen, who passed away unexpectedly in 2017. Edgar was part of the SoLid collaboration since its inception and his efforts were vital to get the experiment started. He will be duly missed.}
}

\arxivnumber{1802.02884} 

\begin{document}
\maketitle
\flushbottom
\section{Introduction}
Significant deficits in the integrated flux of electron anti-neutrinos with respect to theoretical predictions at short distances from fission reactor cores have been reported since 2011~\cite{PhysRevD.83.073006} and are confirmed by recent measurements~\cite{An:2015nua}, together with spectral features in the measured electron anti-neutrino energy spectrum~\cite{RENO:2015ksa, An:2017osx, Abe:2014bwa} that are currently not accounted for by the most up-to-date reactor flux calculations~\cite{Huber:2011wv, Mueller:2011nm, Hayes:2015yka, Huber:2016fkt, Huber:2016xis}. These reactor flux deficits and spectral features, combined with other unresolved anomalies observed in beta-decays~\cite{Kaether:2010ag,Abdurashitov:1998ne,Abdurashitov:2005tb,Giunti:2010zu} and at short baseline accelerator experiments~\cite{Aguilar:2001ty,Aguilar-Arevalo:2013pmq} can be interpreted in various ways, including in terms of a flavour oscillation from the electron anti-neutrino state to one or several new sterile neutrino states~\cite{Gariazzo:2017fdh, Dentler:2017tkw, Berryman:2018jxt} that are not included in the Standard Model of particle physics. 
The inconclusive interpretations based on the current world neutrino oscillation data justify an active program of new short baseline oscillation searches near reactors and accelerators. Some of these are already taking data~\cite{Danilov:2013caa, Ko:2016owz, Serebrov:2017ivv}, while others are currently being commissioned, or under construction~\cite{Minotti:2017iri, Ashenfelter:2015uxt}.
	
The SoLid experiment, short for ``Search for oscillation with a $^6$Li detector'', is a new generation neutrino detector, based on plastic scintillator technology, currently in operation at the BR2 research reactor at the SCK\raisebox{-0.8ex}{\scalebox{2.8}{$\cdot$}}CEN in Mol, Belgium. The experiment will  perform an oscillometric measurement of electron anti-neutrino disappearance as function of the anti-neutrino energy and interaction distance, at a baseline between 6 and 9 meter. The SoLid data will be interpreted in terms of additional sterile neutrinos related to mass eigenstates corresponding to $\Delta m^2\sim 1\,$eV$^2$. Additionally, the experiment will provide a reference energy spectrum for reactor electron anti-neutrinos that are produced from nearly pure $^{235}$U.
The detector design is different from other experiments due to its high level of segmentation, in combination with the use of $^6$Li as an active substance for neutron detection. The BR2 reactor is characterized by its compact core, with an effective diameter of $d_{eff}\simeq 0.5$$\,$m, a thermal power range of 40 -- 80 MW$_{\mathrm{Th}}$, and a fuel matrix that is 93.5\% enriched in $^{235}$U. 

In a previous paper~\cite{Abreu:2017bpe} we outlined the R\&D efforts related to the detector concept and its technology. In this paper we provide details on the construction and operation of a full scale prototype detector module with a fiducial mass of 288$\,$kg. It was deployed near the BR2 reactor in 2015, where it collected a small data sample during reactor operations at a nominal power of 60 MW$_{\mathrm{Th}}$, followed by a longer background measurement campaign when the BR2 reactor was shut down for an extensive overhaul of its Beryllium core matrix. During the latter period, several gamma and neutron sources were also used to investigate the detector response. The main purpose of the prototype experiment is to demonstrate the scalability of the core technology, the stability of operation at the reactor site, the capability to equalize the response of a large number of readout channels in the detector and to perform an initial analysis based on pulse shape discrimination, muon tracking and time correlation of signals. Based on these results, the SoLid collaboration optimized the final design of the 1.6 ton detector, which is currently in operation near the BR2 reactor, with the prime goal of performing an oscillation measurement. 

The structure of this paper is the following: in section~\ref{sec:detector} the detection principle, the chosen technology and the read-out system are described. Section~\ref{sec:br2} discusses the main features of the BR2 reactor and the data taking periods. The event reconstruction and higher-level object definitions are discussed in section~\ref{sec:evreco}, followed by a demonstration of the muon tracking capability of the experiment and a validation of the modelling of the reactor building in combination with state-of-the-art cosmic shower Monte Carlo models in section~\ref{sec:background}. The analysis strategy to detect anti-neutrino events and to reduce the most important backgrounds is outlined in section~\ref{sec:coincidence}, followed by results based on reactor data. We conclude in section~\ref{sec:corrbg} with a discussion on time correlated backgrounds related to decays of trace elements of $^{214}$Bi in the detector components.

\section{Detector description}
\label{sec:detector}

\begin{table}[!t]
	\begin{center}
		\caption{Masses and hydrogen content of the major prototype detector components.}
		\label{tab:protoncounts}
		\begin{tabular}{ccc}
			\hline
			\hline
			Component & Mass (g) & H content ($\times10^{27}$) \\
			\hline
			PVT cubes & $286561.8\pm0.5$ & $14.4822\pm0.0280$ \\
			Tyvek wrappers & $3154.2 \pm 2.4$ & $0.2712\pm0.0055$ \\
			HDPE500 & $44055.0 \pm 73.5$ & $3.787\pm0.007$ \\
			\hline
			Total & $333771.0 \pm 73.5$ & $18.5404\pm0.0293$ \\
			\hline
			\hline
		\end{tabular}
	\end{center}
\end{table}

\noindent 
The inverse beta decay (IBD) process is commonly the most exploited reaction for detecting electron anti-neutrinos with energies in the MeV range:

\begin{equation}\label{eq:beta}
\bar{\nu}_e +p \rightarrow n + e^+,
\end{equation}

\noindent 
which has a threshold energy of 1.806 MeV. The cross-section for the interaction increases with energy, once above this threshold, although this is compensated by the falling energy distribution of anti-neutrinos emitted by the reactor, giving an energy spectrum for the detected anti-neutrinos covering the range from 1.805 to 10$\,$MeV that peaks at approximately 3.5$\,$MeV \cite{Mueller:2011nm}.
The SoLid detector is capable of detecting both the resulting neutron and positron by using a composite scintillation technology. 
\subsection{The prototype detector}
The fiducial mass of the detector is divided in cubical detection cells of dimension $5\times 5\times 5$$\,$cm$^3$. The body of the cubes is made of ELJEN Technology EJ-200 polyvinyl toluene (PVT) based 
plastic scintillator, covered by a  225$\, \mu$m thick $^6$LiF:ZnS(Ag) neutron detection screen from SCINTACOR. The positron deposits its energy directly in the PVT and finally annihilates with an electron, producing a prompt scintillation signal of which the visible energy is directly proportional to the incoming neutrino energy. An electron with an energy of 1 MeV, with a mean path length of $O($mm$)$ in PVT, will typically produce 10000 photons in the scintillator.
Neutrons will predominantly lose energy via elastic collisions until they reach thermal energies and are captured by the $^{6}$Li nuclei in the detection screen. The lower absorption cross section for thermal neutrons of $^{6}$Li compared to other substances such as $^{10}$B or $^{157}$Gd is partly compensated by the large Q-value of 4.78 MeV of the capture reaction:

\begin{equation}\label{eq:LiCap}
^{6}_{3}Li + n \rightarrow ~ ^{3}_{1}H + \alpha
\end{equation}

\begin{figure}[!th]
	\centering
	\includegraphics[width=1\linewidth]{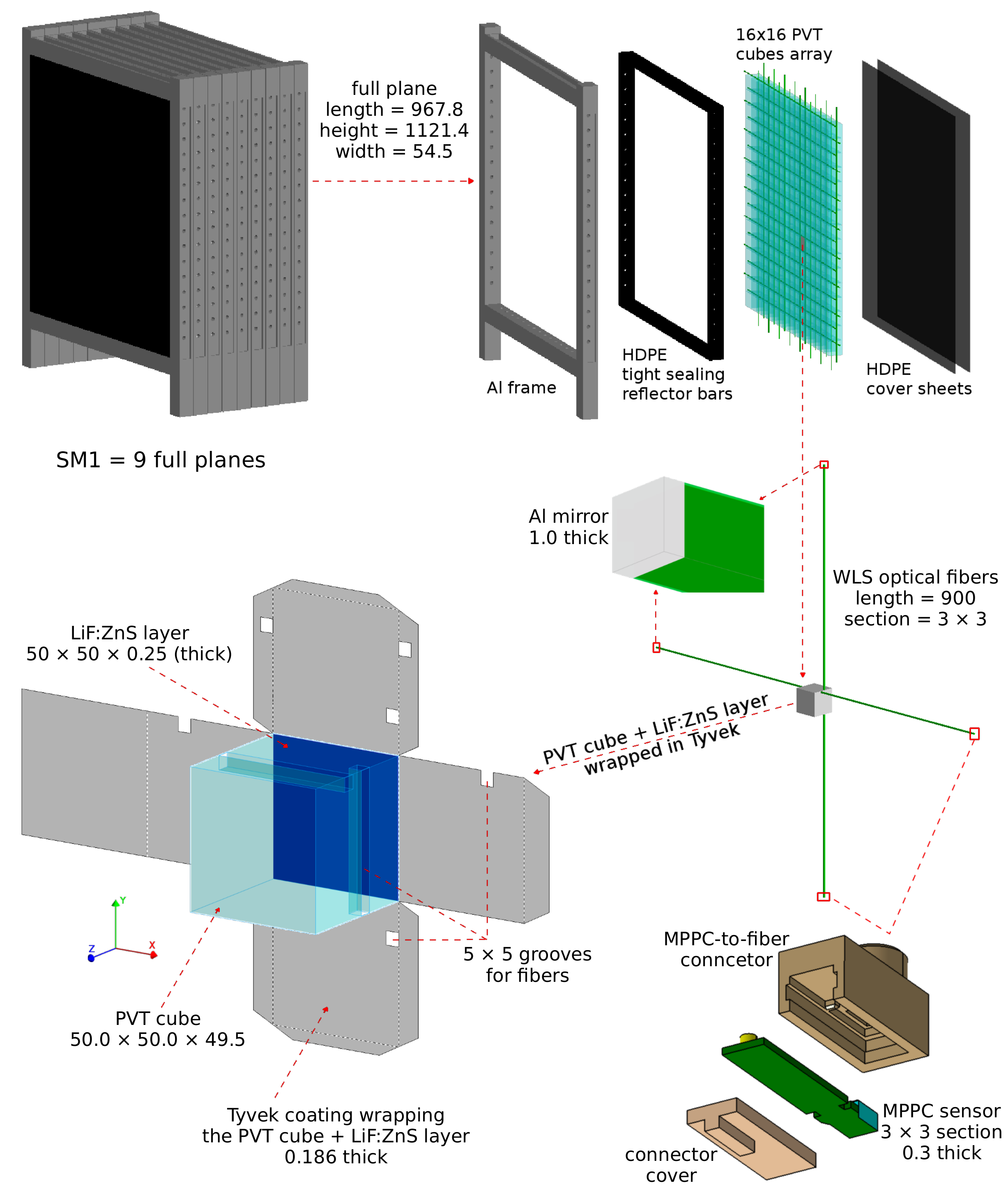}
	\caption{Diagram of the prototype detector, exploded frame, fibre readout and cube assembly. All indicated  sizes are in mm.}
	\label{fig:det_module}
\end{figure}

\noindent
resulting in a large average energy deposit and light yield in the surrounding ZnS scintillator. The time and amplitude structure of the scintillation signals for neutrons and positrons allow for a clear distinction between the two signal components, while the average time difference between the two is characteristic for the thermalisation and capture time of the neutrons travelling through the PVT. The ZnS scintillator is optically coupled to the PVT cube via a small air gap between the two materials, allowing the scintillation light to propagate to the PVT. Each cubic detection cell is optically isolated from its neighbours via a DuPont Tyvek wrapping with an average thickness of 75$\,$g/m$^2$. The light produced either by the ZnS or the PVT scintillators is optically trapped in two wavelength shifting fibres of type BC-91A from St. Gobain, consisting of a core surrounded by a single cladding. These fibres are $3\times3$$\,$mm$^2$ in cross section and are aligned along two perpendicular faces of each cube, in a dedicated groove of $5\times 5$$\,$mm$^2$. All cubes are finally stacked in a $16 \times 16$ cube configuration, composing a detection plane. Each detection plane is lined on the inside with 2cm thick black high density polyethylene (HDPE) to improve the moderation and reflection of neutrons created at the edges of the detector. The outer front and back surface of each detection plane is capped with a 2mm thick black HDPE sheet.
The optical fibres protrude from the edges of a detection plane into a hollow aluminium frame where they are coupled on one end to a multi-pixel photon counter (MPPC) from Hamamatsu type S12572-050P using optical grease. The other end of the fibre is mirrored with a thin aluminium tape. The position of the MPPC and mirror alternates between adjacent fibres to ensure a more uniform light response throughout the detector. The first full scale prototype submodule of SoLid, named SM1 hereafter, consists of 9 collated detection planes to amount to a final configuration of 16$\times$16$\times$9 detection cubes, read out by a network of $32 \times 9$ fibres connected to one MPPC each. The whole module is surrounded by a passive shielding of 9$\,$cm thick HDPE and placed on a steel table for positioning and alignment.

The total proton content of the detector is determined from the average density and weight of all basic components used for construction of the sensitive volume and summarized in Tab.~\ref{tab:protoncounts}.
A schematic view of the main components of SM1 is shown in Fig.~\ref{fig:det_module}.

\subsection{The readout system}
\label{sec:readout}
\begin{figure}[!b]
	\centering
	\includegraphics[width=0.7\linewidth]{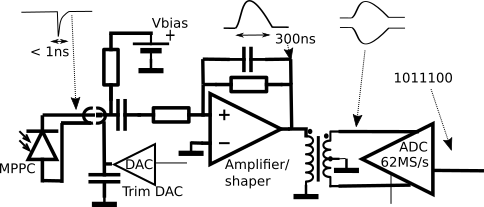}
	\caption{Diagram  of the front-end part of the read-out system.}
	\label{fig:diagram}
\end{figure}

Each MPPC is built up from 3600 pixels, arranged in a 3$\times$3$\,$mm$^2$ array. Each pixel is an avalanche photo diode, and all pixels are connected in parallel. The signal of a discharge of a single pixel is called a pixel avalanche (PA). The read-out system coupled to the MPPCs is designed using custom analogue front-end (AFE) boards, coupled to a front-end trigger board with an FPGA. Triggered data are transferred over a Gigabit network to a DAQ and data storage server. A slow control system is deployed on a separate computer. An overview of the AFE design is shown in Fig.~\ref{fig:diagram}.

\begin{figure}[!b]
    \centering
    \includegraphics[width=0.7\linewidth]{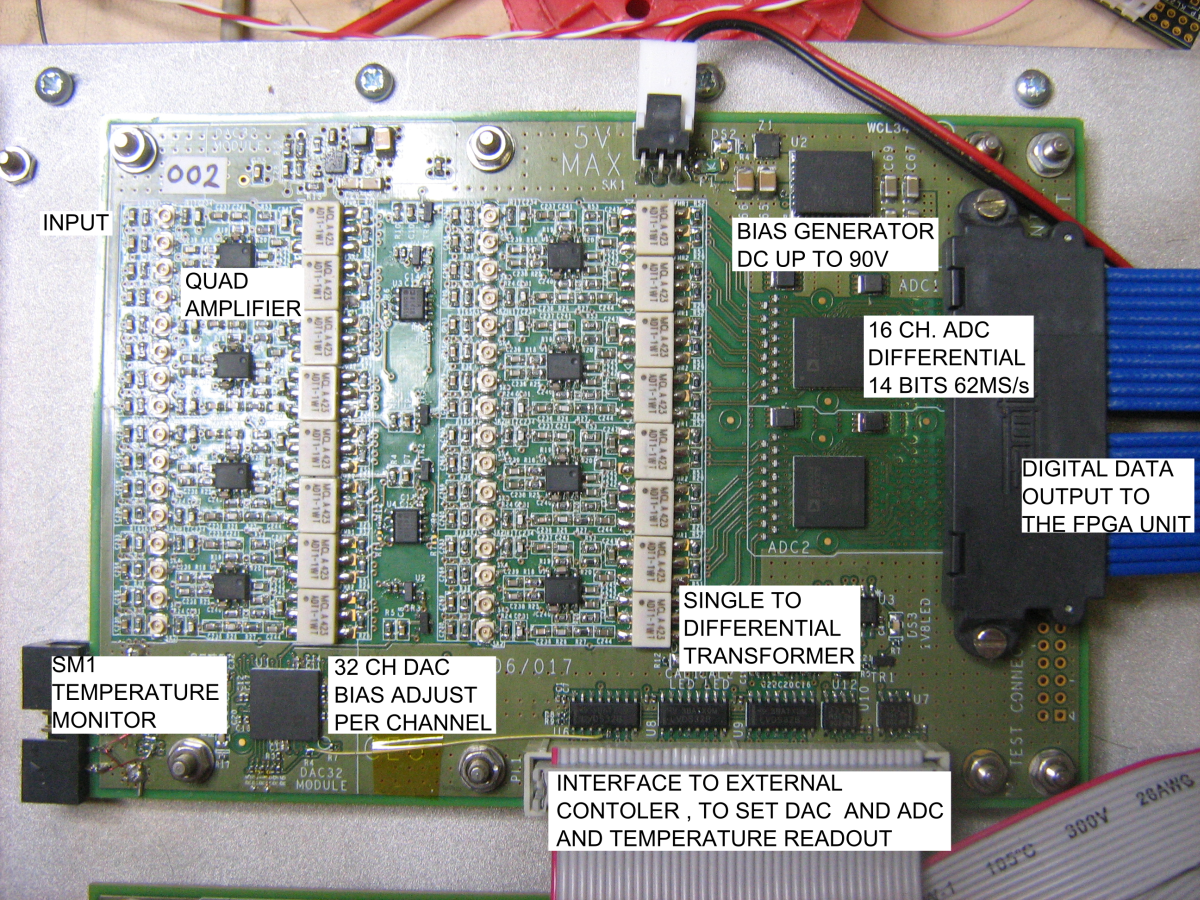}
    \caption{The front-end board, two of which are connected to a readout plane. The most relevant components are indicated on the photograph.}
    \label{fig:elec}
\end{figure}

The signals from each detection plane, containing 32 MPPCs, are collected by a single front-end board as shown in Fig.~\ref{fig:elec} to amplify and digitize the PA signals. The boards also provide a programmable bias voltage for each individual MPPC via a 90$\,$V power supply coupled to the cathode of all MPPCs on the board. The per-sensor bias voltage is then adjusted by providing a programmable low voltage, in the 0 -- 5$\,$V range to the anode of the MPPC. Each MPPC is biased above its breakdown voltage, bringing the pixels into Geiger mode. Since each sensor has an individual breakdown voltage, the bias voltage applied to each sensor has to be individually set to ensure all sensors are operated with a similar bias voltage and thus have a similar photon detection efficiency. 
In this application it is important to keep the dark count and cross talk low. At an ambient temperature of 25$^\circ$$\,$C and at a bias voltage of 1.5 V above the breakdown voltage of the MPPC, the dark count rate, defined as the signal rate with an amplitude above 0.5 PA, is of order 250 kHz and rises strongly with increasing temperature. The cross talk probability, defined as the rate of dark counts with two or more fired pixels, divided by the total dark-count rate, was measured to be 18\% and is fairly independent of the temperature. Reducing the ambient temperature to 5$^\circ$$\,$C will reduce the dark count rate to levels below 100kHz. A precise MC model of the MPPCs used in this phase of the experiment shows that crosstalk values below 20\% do not affect the energy resolution.

The measured light yields are corrected for this effect.
The average light yield per cube was measured in situ via crossing muons and amounted to 12 PA per fibre and per MeV of energy deposited in the PVT of the same cube. This results in an energy resolution of 20\% for electrons and positrons depositing 1 MeV of energy in a single PVT cube. The light yield per cubic cell is also affected by the attenuation of light inside the fibre before it reaches the MPPC. The attenuation length of the fibres  was determined to be 130 $\pm$ 10 cm, using muons crossing the detector. These results are confirmed by measurements using a dedicated laboratory set-up with a $^{207}$Bi source emitting conversion electrons with an average energy of 995 keV~\cite{Abreu:2017bpe}. 

For the full scale detector, several improvements will be adopted to increase the light yield in order to improve the energy resolution. These improvements consist of doubling the amount of wavelength shifting fibres, using double cladded fibres, using thicker Tyvek wrapping, fibre mirrors with a better reflectivity, and a smoother polishing of the cube surfaces. Dedicated laboratory experiments indicate that using these improvements an energy resolution of 14\% for electrons and positrons depositing 1 MeV of energy in a single PVT cube can be achieved. The details of these measurements will be reported in a future paper.

The MPPC signal currents are amplified, relative to a virtual earth, by a LMH6626 low-noise charge integrating operational amplifier, and then
converted into differential signals using a radio frequency transformer, for coupling into the AD9249 analogue to digital converter (ADC). Each analogue board contains two 14-bit ADC chips with 16 channels that have a 2$\,$V peak-peak range, sampling at a rate of 62.5$\,$MS/s, corresponding to time samples separated by 16$\,$ns.
The ADC provides a serialised LVDS output per channel. 
The 16 channel LVDS lines, as well as a number of clock and frame lines are connected to the FPGA boards via a high speed twin-axial ribbon cable. The front-end logic in the read-out system was provided by a single Gigabit Link Interface Board (GLIB)~\cite{Vichoudis:2010voh} equipped with a Xilinx Virtex-6 FPGA controlling two detection planes at a time.

The data trigger logic is implemented in the FPGA of each GLIB board and is based on the time coincidence of PA pulses with an amplitude exceeding a pre-set threshold. This threshold was set to 6.5 PA in order to maintain a sustainable data rate. The time coincidence was imposed by requiring at least one horizontal and one vertical fibre pulse to exceed the amplitude threshold within a coincidence window of 3 samples. When a trigger occurs, a 256 sample waveform is stored from each channel that exceeds the threshold. Additionally, a random periodic trigger was used to read out waveforms from all detector channels.   

Since each MPPC has a unique breakdown voltage, each channel is biased individually to achieve a uniform energy response across the detector. To achieve this, the gain of each channel is measured by identifying the first and second pixel avalanche peaks in the amplitude spectrum of each channel, collected with random triggers. The gain increases linearly with the applied bias over the used bias voltage range and is determined for each channel via a bias voltage scan. During commissioning of the detector, the MPPC gains were equalized to a spread of 20$\,\%$. The remaining variations are equalised off-line using crossing muons to achieve a final RMS of 3\%.

\begin{figure}
	\centering
	\includegraphics[width=0.9\linewidth]{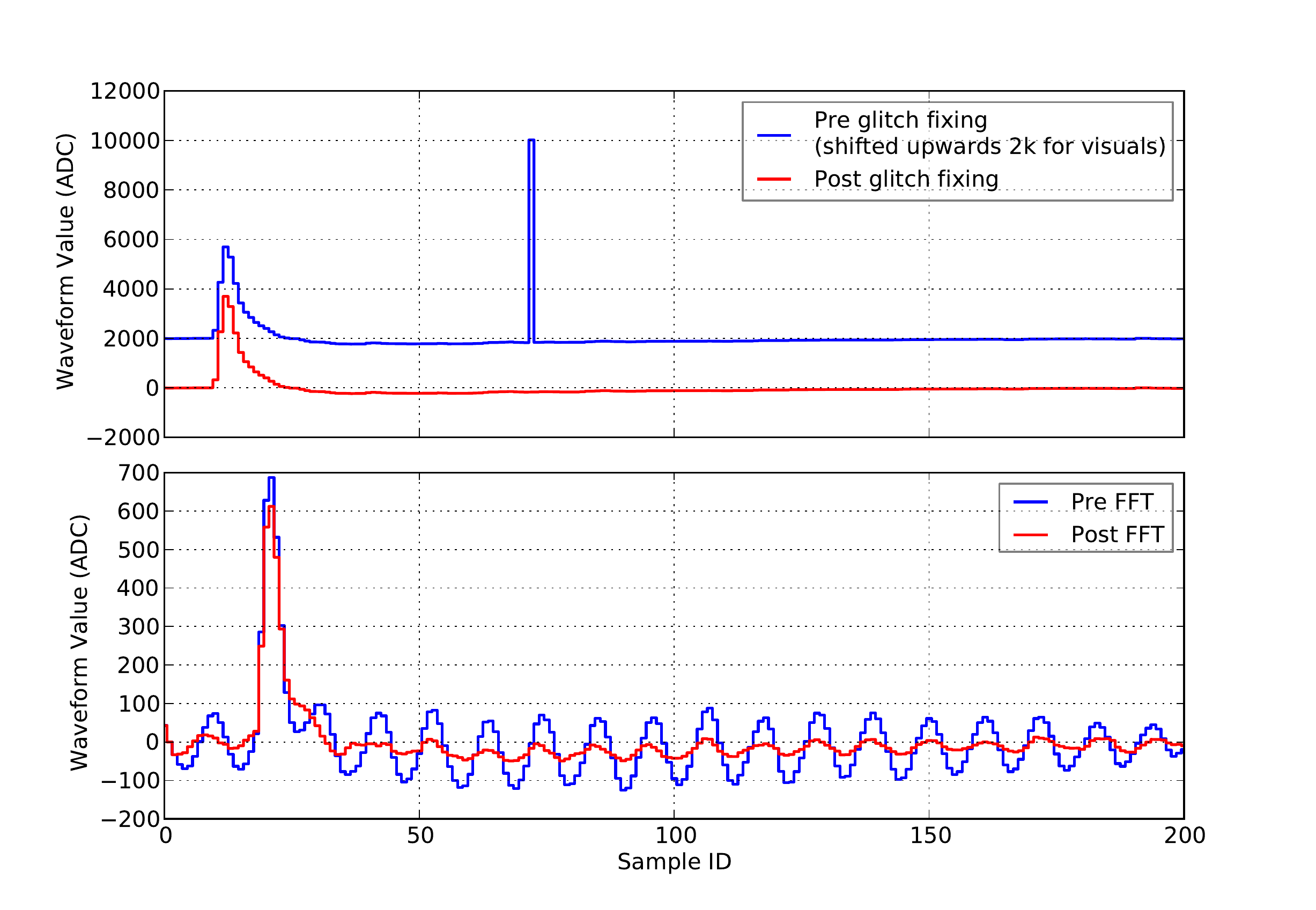}
	\caption{Illustration of LVDS line errors induced in the digitized waveforms (top) and oscillatory noise patterns (bottom). The red curves illustrate the waveform after off-line corrections.}
	\label{fig:noiseExamples}
\end{figure}

\begin{figure}[!th]
	\centering
	\includegraphics[width=0.7\linewidth]{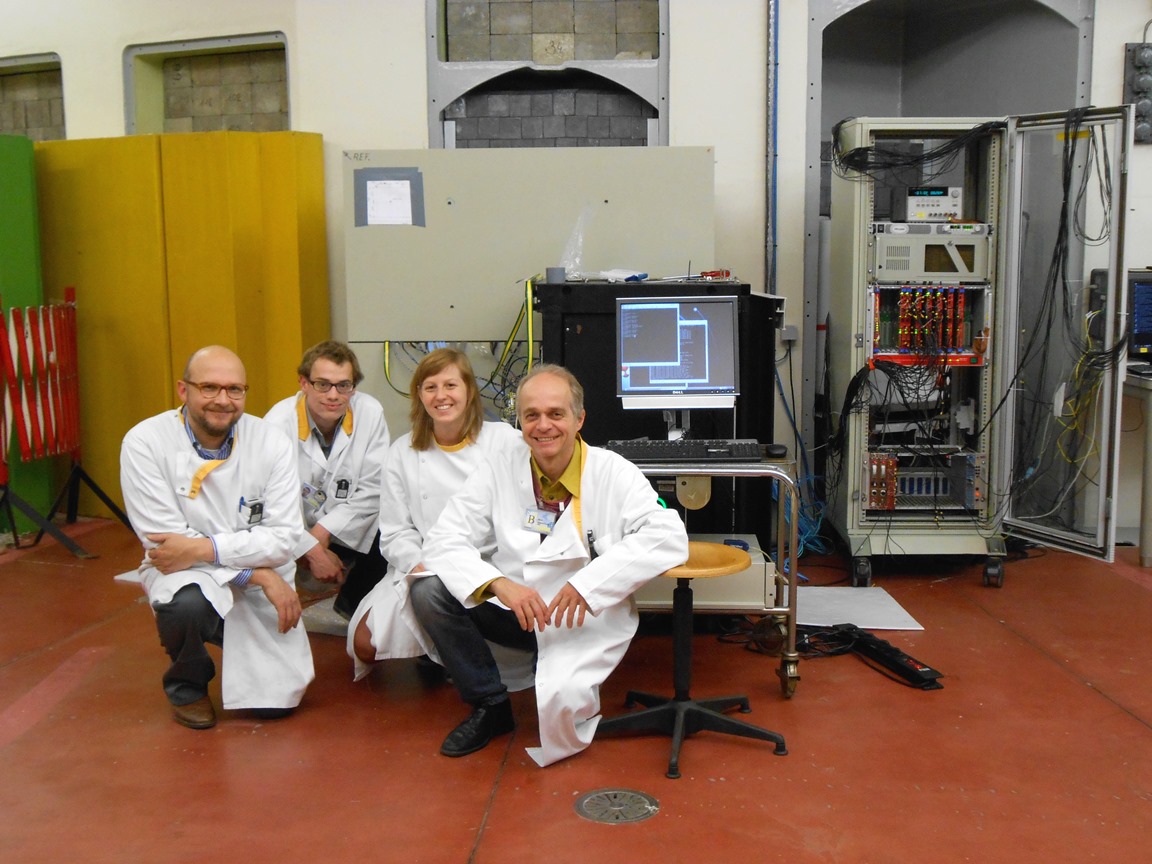}
	\caption{Picture of the prototype deployed at the BR2 research reactor.}
	\label{fig:BR2}
\end{figure}

Due to limited development and construction time for this prototype module, the readout electronics suffered a series of flaws that affected the data taking and trigger rates. Most of the induced effects are mitigated on-line or off-line. They are described below in order of increasing importance.

The bias voltage for the MPPCs is provided by an LT3482 DC/DC converter, operated with a switching frequency of 650 kHz. Read-out channels located in close proximity to the chip  experienced a periodic noise matching the DC/DC converter frequency. Since the amplitude of the switching noise was close to the trigger threshold, amplitude triggers were disabled from these channels. Signals collected from fibres corresponding to these channels, amounting to roughly 2\% of all readout channels, were not used for data analysis. 

A second issue occurred in converting the serialised LVDS signals from the ADC in the FPGA which led to discontinuities in the digitised waveforms. These cause 1 -- 3 data samples to be erroneously higher or lower than the surrounding waveform, as can be seen in the top of Fig.~\ref{fig:noiseExamples}. The amplitude of these erroneous samples is in most of the cases close to half the ADC range and can cause unwanted triggers. To prevent these anomalies from triggering the readout, a programmable digital offset of +127 ADC counts is applied in the ADCs to bring the pedestal level above the mid-point of the ADC range. All the discontinuities will therefore be interpreted as negative offsets and do not cause a trigger. The discontinuities are filtered out off-line by a linear interpolation between adjacent sample values. 

The amplification circuit also induced oscillations and a small undershoot in the recorded waveforms, as shown in the bottom of Fig.~\ref{fig:noiseExamples}. The worst impact to the waveforms comes from the beating of oscillations with frequencies between 6 and 8 MHz. The amplitude of the oscillations varies significantly between each amplifier board. In the worst affected boards the maximum amplitude of these oscillations, when in phase with each other, could be of the order of three PA. The oscillating noise required that the trigger threshold was set to 6.5 PA resulting in an average energy cut-off of roughly 500 keV on the triggered neutron waveform amplitudes. Future improvements in noise reduction and triggering will further reduce this threshold.
Under the conditions described here, low amplitude signals occurring in phase with the oscillation minima have a reduced trigger efficiency. The oscillations have a small impact on the reconstructed amplitude of scintillation pulses created inside the PVT. These amplitudes can be corrected by either fitting the oscillation and removing it or by using an FFT (Fast Fourier Transfer) based filtering method to remove the signal components close to the noise frequencies. The final average impact of the noise after corrections can be conservatively estimated to be 2\% on the waveform amplitude and less than 2\% on the waveform integral.

The oscillations had a similar effect on the pulse shape discrimination between neutron and positron/gamma induced signals since the signature of signals from neutron capture in the $^6$LiF:ZnS(Ag) consist of an initial scintillation pulse followed by low amplitude pulses occurring for up to a microsecond after the arrival of the first photons.

The unexpected flaws in the readout electronics induced a serious setback in the trial run with this prototype module with overall lower neutron detection efficiencies and less performant pulse shape discrimination compared to expectations. Nevertheless, relevant physics measurements could be performed in realistic reactor conditions which are described in the following sections of this paper. Based on these results, the electronics design was revisited for the full scale SoLid detector~\cite{1748-0221-12-02-C02012}.
\section{The BR2 reactor at SCK$\cdot$CEN}
\label{sec:br2}
The SoLid experiment is operated at the BR2 reactor of the Belgian Nuclear Research Centre, SCK\raisebox{-0.8ex}{\scalebox{2.8}{$\cdot$}}CEN, in Mol. 
The twisted design of the BR2 core allows for a small effective core diameter of $d_{eff}\simeq 0.5$$\,$m in combination with a considerable thermal power in the range of 40 -- 80 MW$_{\mathrm{Th}}$. Its fuel matrix is 93.5\% enriched in $^{235}$U, with a maximal average burn-up of the enriched fuel of $\sim$ 50\%.
The detector is installed in direct line of sight with the nominal center of the reactor core facing the concrete reactor pool wall at a distance of 6 meter. The detector is shielded with an additional 20 cm lead in between the detector and the reactor pool. No experiments or beam lines surround the experiment on this floor of the reactor containment building. The overburden of the experiment is small ($O(30$$\,$m.w.e.$)$) and not effective in shielding the detector from cosmic rays. 

\begin{table}[!t]
	\begin{center}
		\caption{Summary of data taking periods of the prototype detector. Two calibration runs taken with an AmBe neutron source were taken in front and in the back of the detector, denoted as calibration 1 and 2.}
		\label{tab:datasets}
		\begin{tabular}{ccc}
			\hline
			\hline
			Dataset type & Dates & Live time (hrs) \\
			\hline
			Reactor on & 00:00 21 Feb - 08:00 24 Feb & 50.9 \\
			Reactor off & 08:00 24 Feb - 00:00 12 Mar & 577.8 \\
			&  and 00:00 27 Mar - 12:00 11 Apr  &\\
			Co calibration & 14:00 22 Apr - 14:00 24 Apr & 48 \\
			AmBe calibration Pos.~1 & 15:42 28 Apr - 24:00 30 Apr& 65.0 \\
			AmBe calibration Pos.~2 & 17:00 4 May - 18:00 4 May& 1.0 \\
			\hline
			\hline
		\end{tabular}
	\end{center}
\end{table}

In the winter of 2014-2015 the prototype detector was installed in front of the reactor core as shown in Fig.~\ref{fig:BR2} and commissioned during the last days of a reactor cycle. At the end of February 2015 the reactor was shut down for a 1.5 year-long overhaul of its Beryllium fuel core matrix. During this shut-down period, background data was collected, as well as some source calibration data. A list of data sets taken with the prototype can be found in Tab.~\ref{tab:datasets}.  

\begin{figure}[!t]
	\centering
	\includegraphics[width=0.9\linewidth]{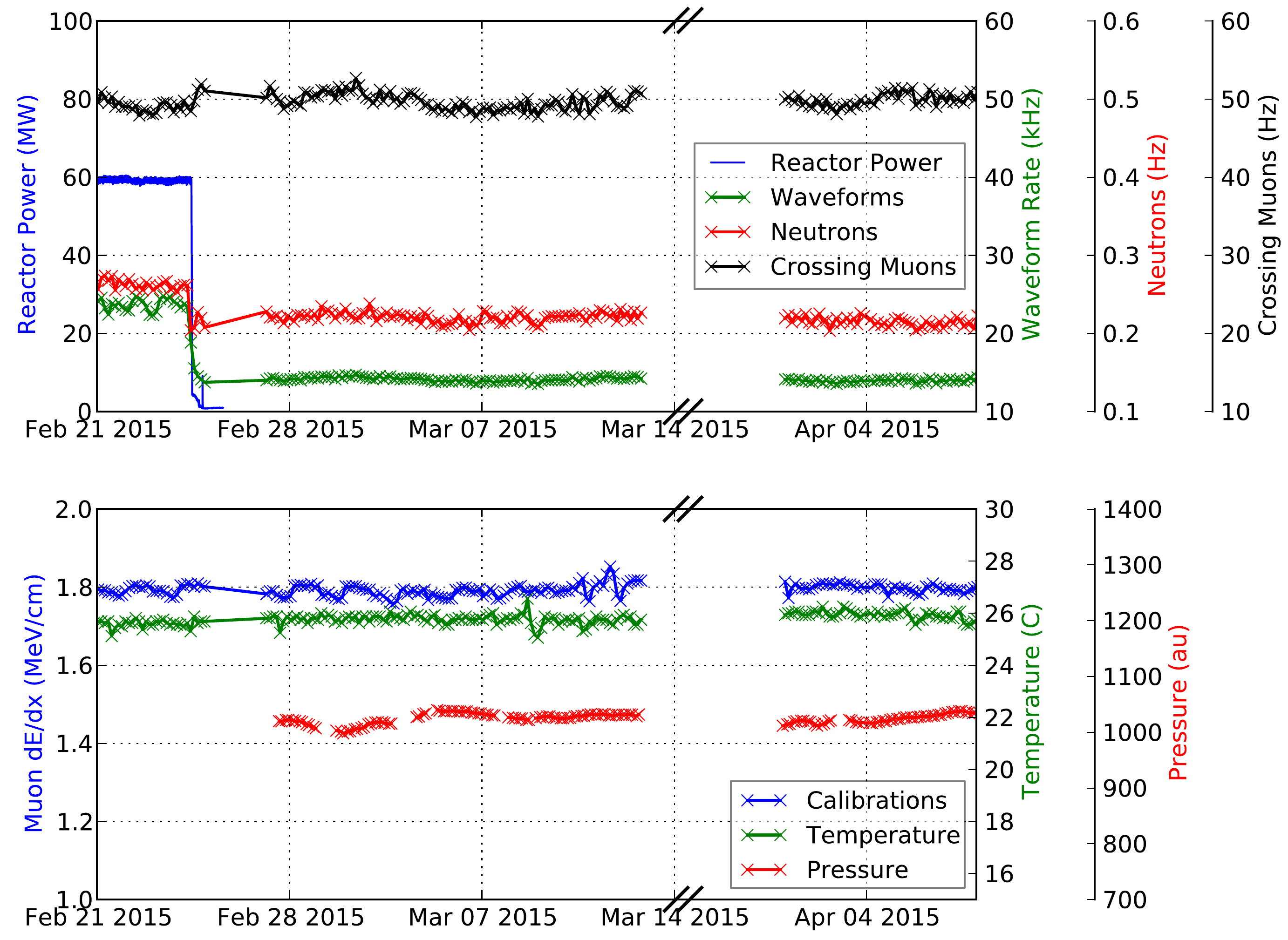}
	\caption{Data trends for prototype data taking. The second half of March has been removed due to various environmental tests conducted in the BR2 building itself.}
	\label{fig:trends}
\end{figure}

Figure~\ref{fig:trends} shows various trends found in the data, for both the reactor on and off data taking periods. The identification of the various objects is explained in later sections. It can be seen that the rate of reconstructed cosmic muons is stable over this period, which indicates that the timing of the detector is stable. The RMS of the deposited energy per path length of reconstructed muons crossing the detector over a period of 3 months was measured to be less than 1\%, proving the stability of the gain and energy response of the detector. The reactor shut-down is indicated in the power profile, which is correlated with the dropping rate of triggered waveforms exceeding the thresholds. The rate of reconstructed neutron signals shows a very small decrease after the reactor shut-down, implying a small reactor induced neutron background. Data taking dead-time was monitored continuously and was found to be negligible. 

\section{Event reconstruction}
\label{sec:evreco}
The time coincidence requirement of the on-line trigger pairs every signal seen on a horizontal fibre with at least one signal on a vertical fibre in the same plane in order to find the cube from which the signal originated. 
In an environment with low neutron rates, the majority of signals correspond to light created by scintillation of the PVT. These PVT signals 
can be induced by electrons, positrons, muons, gammas and proton recoils from fast neutrons. The waveforms of these PVT signals are typically short in time with a pulse length less than 300 ns, and the amplitude is proportional to the energy deposited in each cube.
A cube signal is a combination of a horizontal and a vertical peak, based on the time coincidences seen by the trigger. Time ordered peak lists of horizontal and vertical peaks are compared in a double loop. When a match is found, both peaks are joined into a cube and are labelled as assigned. A peak can only be assigned once. 

Due to the applied calibration constants, discussed in the next paragraphs, the energy in a cube is defined as the sum of the attenuation corrected energy of the two peaks.
Two independent corrections are applied on the peak level: one to take out variations in the light collection efficiency and gains across channels, and one to take out attenuation effects inside the optical fibres. 
\subsection{Energy corrections}
\label{sec:chcorr}

\begin{figure}[h]
	\centering
	\includegraphics[width=0.7\linewidth]{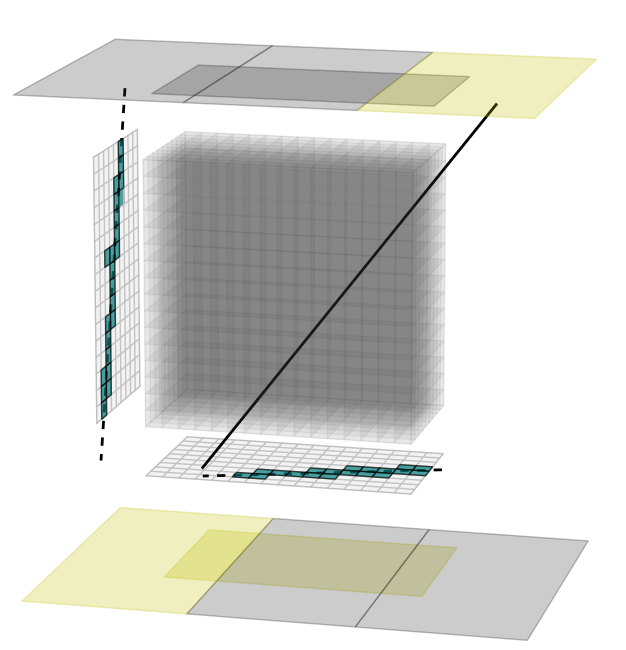}
	\caption{Example muon track. The clusters of triggered channels are highlighted on the two MPPC arrays as blue squares, and the result of the linear regression  to the reconstructed cube positions is shown by the solid line.}
	\label{fig:egTrack}
\end{figure}

Due to the highly segmented structure of the detector, muons can be tracked very easily. Using a straight line, their trajectory through the detector can be fitted.  An example of an event display showing a muon track fit is shown in Fig.~\ref{fig:egTrack}. From this fit, the track length of a muon in each cube is obtained. This allows the integral of the waveform over track length ratio $\left( I/x \right)$ to be calculated for each individual channel and for the average across all channels in the detector. Each channel is equalized w.r.t. the detector average by multiplying its energy response $\left( I/x \right)_i$ with a correction factor $\delta_i$:

\begin{equation}
\delta_i = \frac{\left( I/x \right)_{det}}{\left( I/x \right)_i},
\end{equation}

where $\left( I/x \right)_{det}$ corresponds to the detector average. This step reduces the RMS of the channel responses, $\left( I/x \right)_i$, from $43\,\%$ to $3\,\%$.

\begin{figure}
	\centering
	\includegraphics[width=0.6\textwidth]{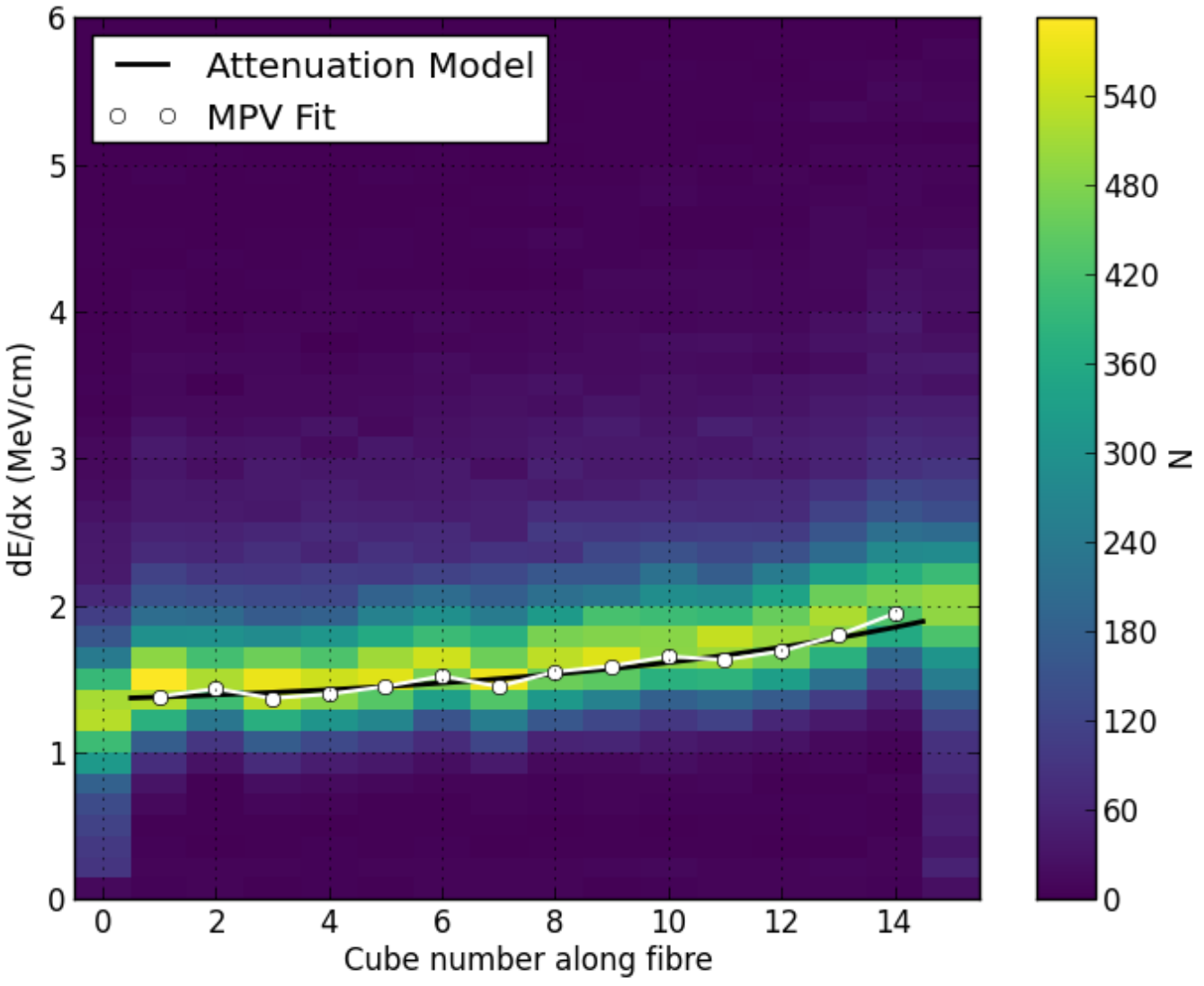}
	\caption{The observed muon $dE/dX$ in the detector, corrected for channel gain variations and offset with respect to the expected value for PVT, as a function of cube - MPPC distance (histogram and white dots) compared to the predicted values from first principles (black).}
	\label{fig:att}
\end{figure}

In a second step, the conversion between $I/x$ and $dE/dx$, the deposited energy per unit track length, is made by comparing the reconstructed energy per path length with the theoretical $dE/dx$ for a minimally ionising muon in PVT, which is $1.776\,$$\,$MeV/cm. The reconstructed energy in each cube, given by the sum of the corrected energies in both channels yielding a cube signal, is therefore multiplied with the ratio of the theoretical $dE/dx$ for muons and the mean $I/x$ across the detector. The corrected $dE/dx$ distribution for reconstructed muon hits in the detector is shown in Fig.~\ref{fig:att} as a function of the distance of the cube w.r.t. the MPPCs that contributed to the signal. The values show a clear dependence on the distance the light travels to reach the MPPC. This attenuation of light in the fibre can be parametrized and is taken into account in the final energy reconstruction by means of a simple light loss model:

\begin{equation}
F_{loss}(d) = \frac{1}{2}\left( e^{-d/a} + re^{(2X - d)/a} \right)
\label{eq:att}
\end{equation}

with X the total fibre length of 80$\,$cm. 
When this model is fitted to $dE/dx$ data, the attenuation constant $a$ and the reflectivity of the mirror $r$ are determined to be respectively $1.3 \pm 0.1\,$m$^{-1}$ and $0.85 \pm 0.03$. Due to being in a different optical environment, the cube closest and furthest from the MPPC are excluded from the fit. All edge cubes have at least one face touching the black HDPE, reducing the reflectivity compared to a cube that neighbours four white Tyvek wrappings. Once the position of the cube has been reconstructed, the reconstructed energy is corrected to a zero loss value using equation \ref{eq:att}.

\subsection{Pulse shape discrimination}
\label{sec:NeutronId}
\begin{figure}[t]
\centering
\includegraphics[width=0.6\textwidth]{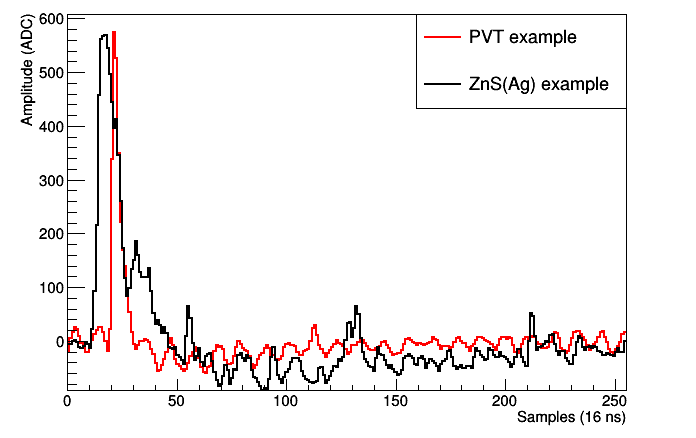}
\caption{Comparison of an amplified and digitized PVT signal (red) and an equivalent ZnS(Ag) signal. Both pulses are selected to have similar amplitudes for better comparison.}
\label{fig:PVTvZnS}
\end{figure}

Determining if an isolated cube signal is a  signal induced by a neutron interacting with the  $^6$LiF:Zns(Ag) is done by exploiting the difference in time constants of the two scintillating materials found within a detection unit. The small time constant of PVT means energy deposited is released almost instantaneous, resulting in a single sharp peak. In contrast, ZnS(Ag) has a much larger time constant. Energy deposited in the $^6$LiF:Zns(Ag) is released over the course of several microseconds, giving rise to several peaks in close succession. An example of a PVT and a ZnS(Ag) signal with similar amplitude can be seen in Fig.~\ref{fig:PVTvZnS}. In this figure there is some periodic noise and undershoot present in the waveforms, these features were discussed in section~\ref{sec:readout}. 

This difference in pulse shape is exploited by using a simple integral over amplitude ratio, $\mu=I/A$. An integral range of 40 time samples is chosen such that it is an integral number of times the main noise period (8 samples) and it maximises the discrimination power between PVT and ZnS(Ag) signals. 

Channel dependency in the integral value is equalized in a similar way as the energy response. The $\left( \mu_i \right)$ value of each channel is corrected by the ratio of its mean value w.r.t. the median of all values across the detector, obtained from a neutron poor sample of background data during periods when the reactor is off.


\begin{figure}
\centering
\includegraphics[width=0.75\textwidth]{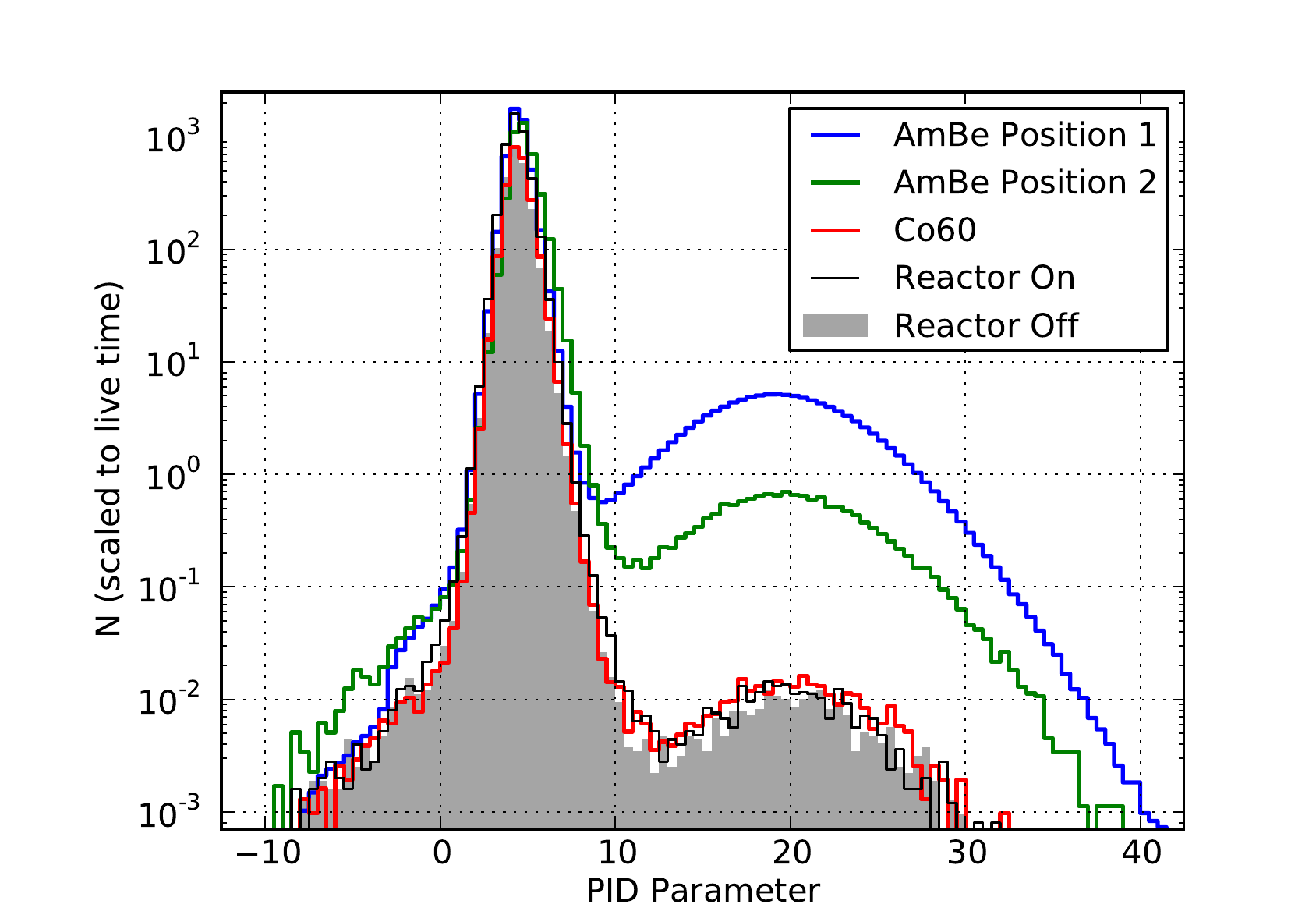}
\caption{The combined neutron discriminant plot for several data sets that are either depleted in neutrons (filled histogram, black and red line), or with an increased neutron rate (green and blue lines).}
\label{fig:nPId}
\end{figure}

To further boost the discrimination capabilities of the corrected integral over amplitude ratio, information from the two channels contributing to the cube is added to obtain

\begin{equation}
PID = \mu_x \times cc_x + \mu_y \times cc_y,
\end{equation} 

where $cc_{x,y}$ are the corresponding correction factors for each of the two channels contributing to the signal. The distribution of the PID value for event samples that are either depleted or enriched in neutrons is shown in Fig.~\ref{fig:nPId}, indicating that neutron induced signals correspond to high values of the PID discriminant, while scintillation signals created directly in the PVT by electrons, positrons and gammas correspond to low values. Signals created in the PVT, characterized by a PID value smaller than 10 will be referred to as electromagnetic or EM signals in what follows. Neutron-like signals are characterized by a PID value larger than 10, and yield a high identification efficiency of 98 $\pm$ 1\% on triggered waveforms. 

\subsection{Neutron detection efficiency}
\label{sec:neutroneff}

\begin{figure}[!t]
	\centering
	\includegraphics[width=0.65\linewidth]{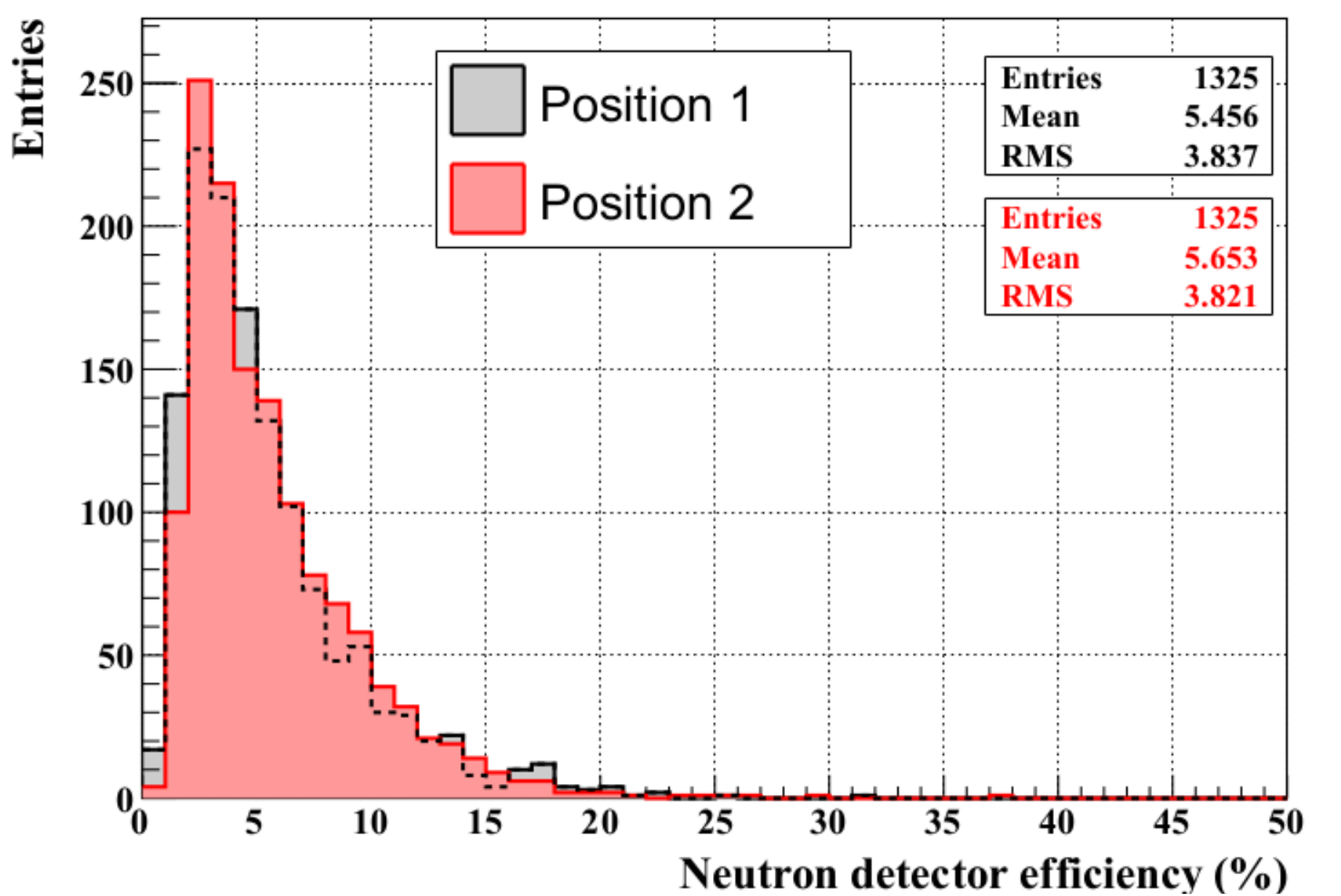} 
	\caption{The neutron detector efficiency component, $\epsilon_{det}$, as determined with an AmBe calibration source, on a cube-by-cube basis for two different source positions). }
	\label{fig:e_det_d}
\end{figure}

The detection of electron anti-neutrino interactions via the IBD process requires an efficient neutron detection. The neutron detection efficiency, $\epsilon_n$, can be factorized in two parts: 

\begin{align}
\epsilon_n = \epsilon_{Li} \times \epsilon_{det},
\label{eq:eff}
\end{align}

with $\epsilon_{Li}$ the probability that the neutron gets captured on the $^6$LiF:ZnS(Ag) screens, and $\epsilon_{det}$ the subsequent probability that its signal will be identified by the detector trigger and off-line pulse shape discrimination.
As  discussed in an earlier paper~\cite{Abreu:2017bpe}, $\epsilon_{Li}$ was derived through \GEANTfour~\cite{Agostinelli:2002hh} simulations. 
Neutrons were generated with the proper IBD energy and kinematics and $\epsilon_{Li}$ was estimated by the ratio of the neutrons that capture on $^6$Li to the number of generated events. The average value in the whole detector was found to be $\epsilon_{Li} = (  52.01 \pm 0.53\textrm{(stat)}  \pm 3.06\textrm{(syst)}  )\,\%$. Note that the systematics come from the uncertainty on the $^6$Li content of the $^6$LiF:ZnS(Ag) screens. 

\begin{figure}[!t]
	\centering
	\begin{subfigure}[b]{0.45\textwidth}
		\includegraphics[width=\textwidth]{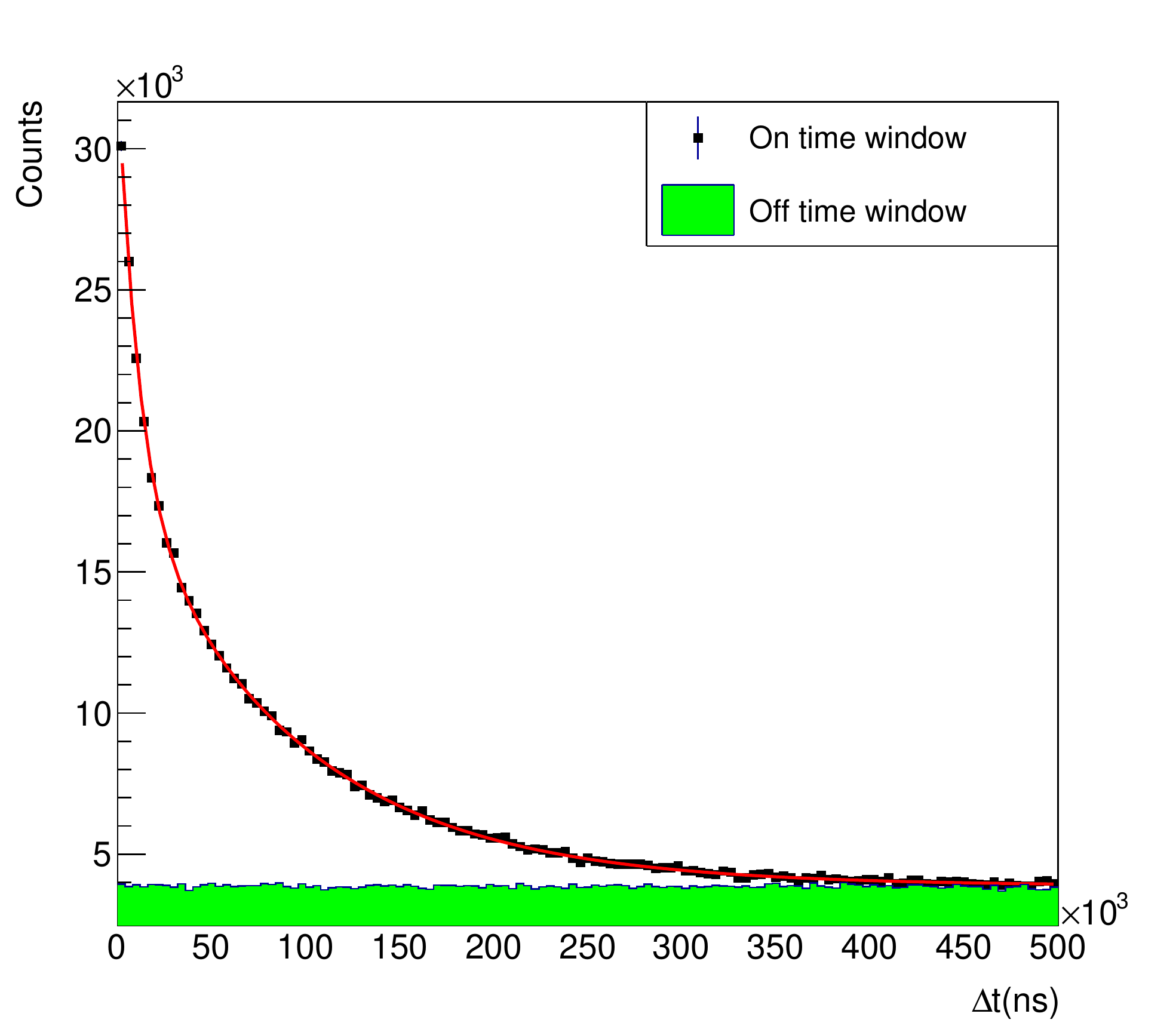}
	\end{subfigure}
	\begin{subfigure}[b]{0.45\textwidth}
		\includegraphics[width=\textwidth]{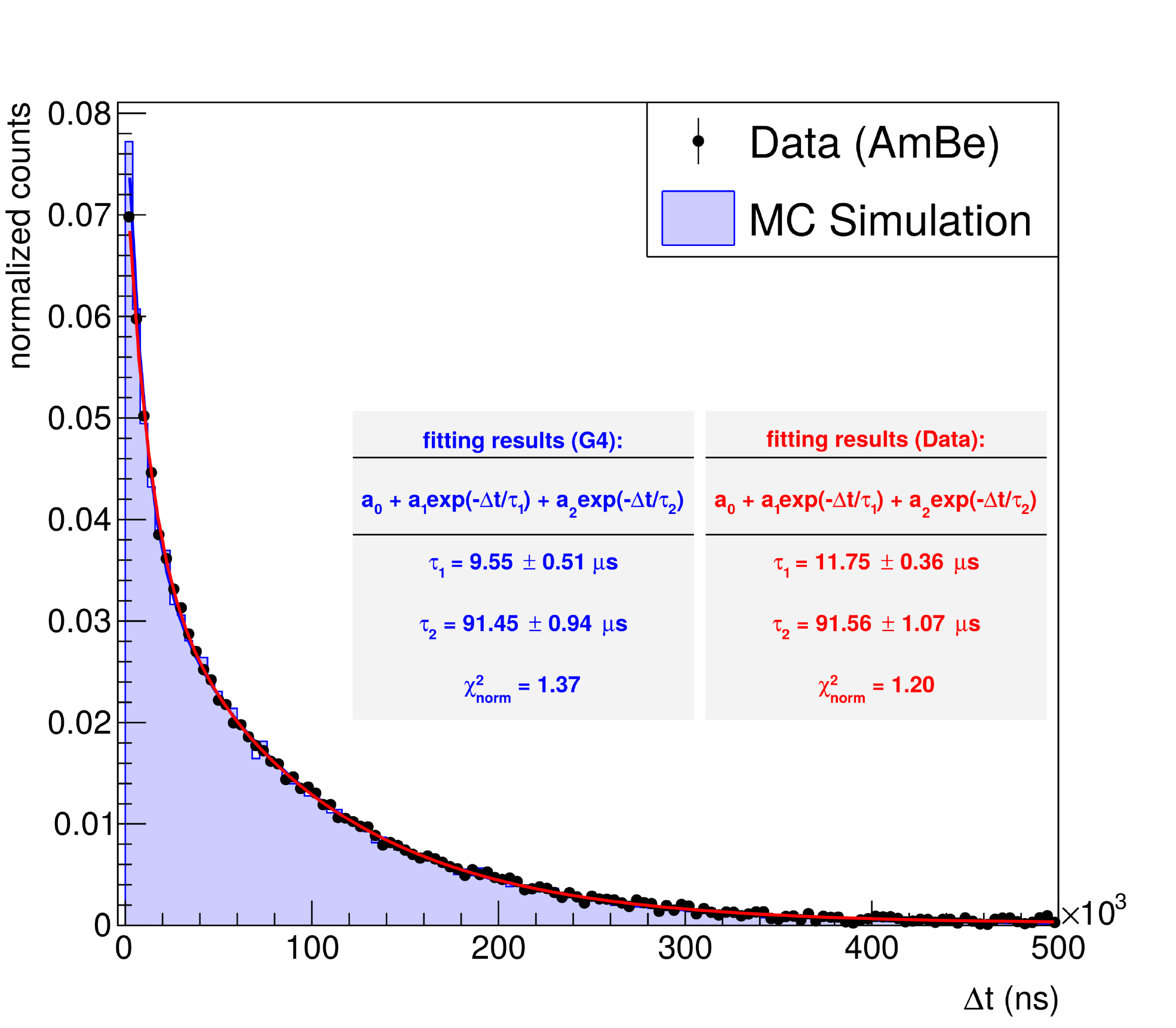}
	\end{subfigure}
	\caption{Distribution of $\Delta$t values between prompts and neutrons for the AmBe data (left) and comparison between \GEANTfour predictions and AmBe data with off-time window determined background subtracted (right).}
	\label{fig:neutrondecay}
\end{figure}

On the other hand, $\epsilon_{det}$ was determined using calibration data with neutrons emitted by an AmBe source, placed either in the front (Position 1) or in the back (Position 2) of the detector, and a neutron selection based on a PID value higher than 10. The values of $\epsilon_{det}$ were taken from the number of neutrons that were identified in each cube, divided by the expectation coming from simulations.  The distribution of all values of $\epsilon_{det}$ obtained for 
cubes receiving a sufficient amount of signal hits during the two calibration runs are shown in Fig.~\ref{fig:e_det_d}. The values obtained from both runs are in good agreement and combined to obtain a final average across the detector of $\epsilon_{det} = (  5.51 \pm 0.02\textrm{(stat)} \pm 1.21\textrm{(syst)}  )\, \%$. The systematic uncertainties come from the uncertainties on the source activity and prediction of the neutron rate in each cube. 
Additionally, uncertainties on the measured neutron rates were employed to account for the uncertainty in the run-time, induced by the dead-time due to the large activity of the AmBe source. 
Note that the systematics  of the measurement ($\sim $20\%) are very conservative owing to this dead-time problem. 
The final number of $\epsilon_{n} = (  2.87 \pm 0.65  )\, \%$ was found combining the average values of $\epsilon_{Li}$ and $\epsilon_{det}$. 
This number gives an approximate value in the detector, and for a more accurate study one would have to combine the  $\epsilon_{Li}$ and $\epsilon_{det}$ values found per cube using dedicated calibration runs. The low neutron detection efficiency is primarily due to the high trigger threshold conditions and the noise conditions affecting the pulse shape discrimination. These results were taken into account in a new design of the signal processing electronics for the next phase of the detector development. The main improvements with respect to neutron detection and triggering involve doubling the amount of neutron detection screens per cube, raising the capture efficiency to 66\%, combined with a new neutron trigger algorithm that is shown to yield efficiencies of $\sim$ 80\%. Both improvements, combined with a similar waveform discrimination will yield a neutron detection efficiency larger than 40\%.

\subsection{Neutron capture time}
\label{sec:neutroncaptime}

The AmBe calibration data is also used to perform a study to estimate the neutron capture time. This is the time that a neutron needs to thermalise and gets captured by a $^{6}$Li atom. A neutron loses energy in the detector by scattering from a nucleus, mostly H and C in the case of SoLid, and in doing so it creates proton recoils. The time correlation between a prompt signal corresponding to a proton recoil and the delayed $^6$Li neutron capture provides us with a characteristic time that is related to the thermalisation and capture processes of the neutrons, shown in Fig.~\ref{fig:neutrondecay}. 
A clean neutron enriched sample was selected from the AmBe calibration runs, and an off-time window is used to model the background in the data. The neutron capture time of (91.56 $\pm$ 1.07)$\,\mu$s is estimated from data, which is in very good agreement with (91.45 $\pm$ 0.94)$\,\mu$s determined from a \GEANTfour simulation.
\section{Cosmic ray events}
\label{sec:background}

The SoLid prototype was installed on surface, with an approximate overburden of 30 meters water equivalent (m.w.e.). Therefore, a significant flux of atmospheric muons and neutrons is expected to interact with the detector. Some of these interactions can create fast neutron signals that look similar to time-correlated IBD events due to the proton recoils observed before thermalisation and capture. These fast neutrons are either part of the incoming cosmic ray flux, or by-products from spallation reactions induced by muons interacting with material inside or surrounding the detector. Slow neutrons can contribute to the accidental backgrounds, where a gamma ray and a slow neutron are detected accidentally  within the time window used to search for IBD signals. 

Cosmic ray events can, on the other hand, also be exploited to calibrate the energy response of the detector, as described in section~\ref{sec:chcorr} and monitor the long-term stability of data taking with the detector. Muon rates and angular distributions are used to validate the cosmic ray models and simulation of the material structures in and around the detector used by the experiment.
Time-delayed signals following a tagged muon validate the timing and pulse shape discrimination based particle identification, and will allow for a data driven estimation of the time-correlated background components. This will in turn allow the development of dedicated selection criteria to reduce the contributions of these backgrounds.

\begin{figure}[!b]
	\centering
	\includegraphics[width=1.0\linewidth]{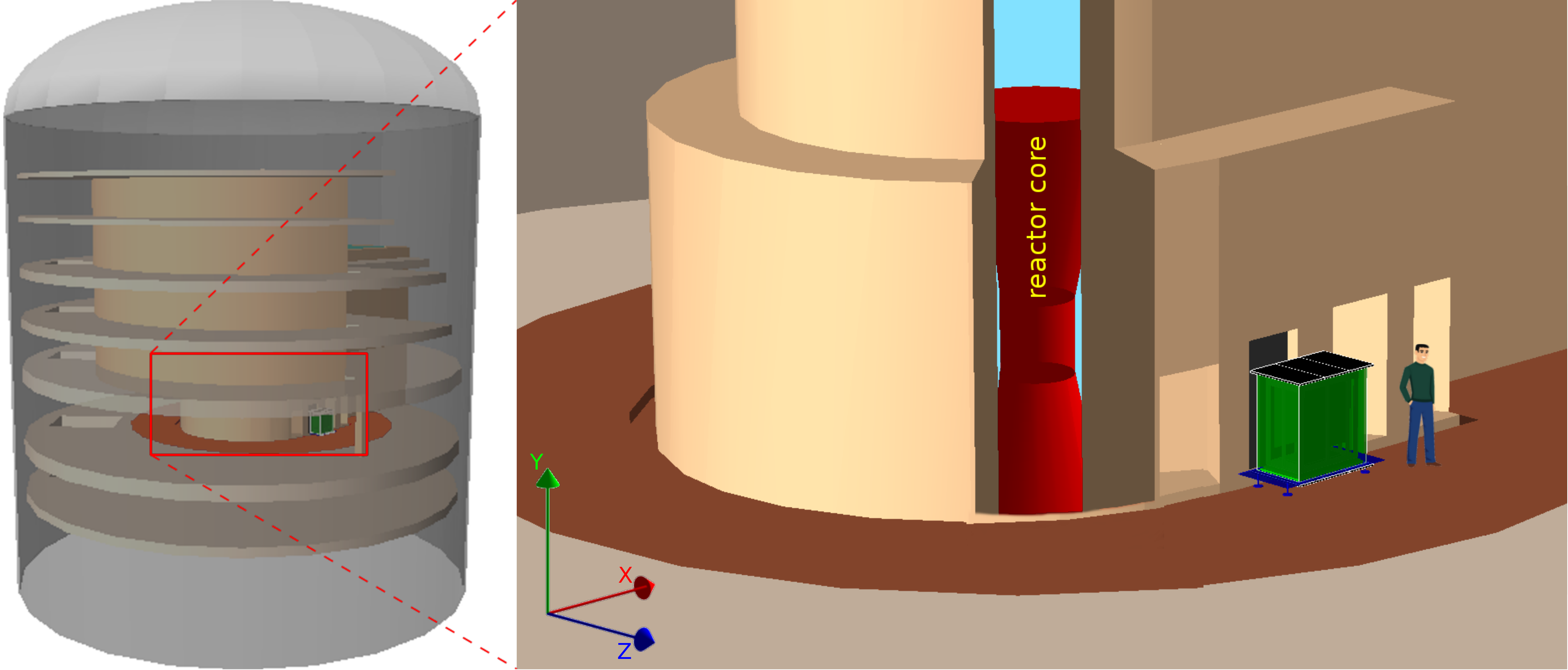}
	\caption{Sketch of the BR2 containment building and the prototype detector simulated in \GEANTfour.}
	\label{fig:geant}
\end{figure}

To estimate muons and muon-induced backgrounds, 
simulations of single muons were generated in a rectangular surface above BR2 using the analytical flux of Guan et al.~\cite{Guan:2015vja}. 
The propagation of muons, through the surrounding material, was performed using \GEANTfour. 
A detailed description of the BR2 containment building and the prototype was input in \GEANTfour, shown in Fig.~\ref{fig:geant}, and the events that hit the detector were processed by a standalone code that simulates the effect of the read-out system. 
On the other hand, atmospheric neutrons were generated using the model of Gordon et al.~\cite{10.1109/TNS.2004.839134}.  
Large samples of muons and atmospheric neutrons were produced using the recipe described above and they were analysed using the SoLid off-line software.  
This procedure allowed us to estimate the flux and energy distributions of muons, spallation and atmospheric neutrons in the prototype.

\subsection{Muon identification \& reconstruction}
\label{sec:MuonID}

Muons can be efficiently tagged given their high channel multiplicity and relatively large energy deposits. Efficient tagging is critical for the neutrino analysis, since muon spallation can lead to increased backgrounds in the IBD analysis. It is also vital to ensure that positrons are not falsely tagged as muons, thus reducing the overall efficiency of detecting neutrinos. 

\begin{figure}[!t]
	\centering
	\includegraphics[width=0.55\linewidth]{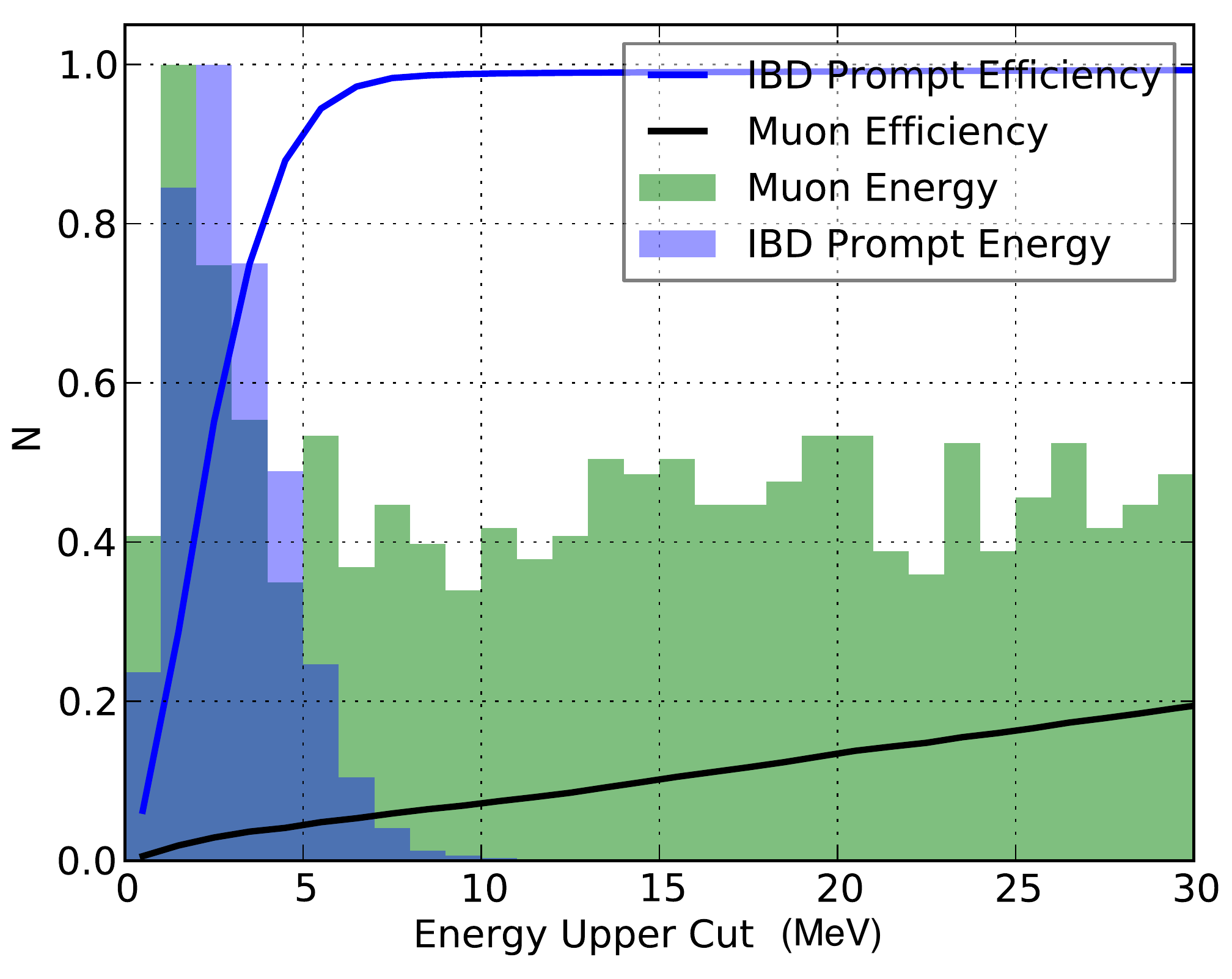}
	\caption{Simulated energy distribution of cosmic muons and IBD positrons (solid histograms). The cumulative distributions for an increasing upper limit in visible energy are indicated by the solid lines.}
	\label{fig:muonPosEnDists}
\end{figure}

The simulated energy deposited in the PVT by muons, and IBD positrons is shown in Fig.~\ref{fig:muonPosEnDists}. Simulation shows that 7$\%$ of muons interacting with the detector, deposit a total energy in the PVT below 8 MeV. This minimum total energy deposit is used as a simple criterium to tag crossing muons. For the remaining low energy muons failing this criterium, it is found that around half of the events involve the outer edge cubes of the detector, which can be used as a veto region.

The segmentation of the detector allows the path of the muon to be estimated in cases where many channels triggered simultaneously. 
An example muon track is shown in Fig.~\ref{fig:egTrack}. The path of the track is parametrised as a straight line, with the $xz$ and $yz$ projections fitted separately using linear regression. Comparisons with simulation have shown that the precision of reconstructing the path-length through a cube, averaged over the detector, when requiring a minimum of 7 hits on both arrays of MPPCs, is around half a centimetre. This hit multiplicity is the default requirement for resolving a muon track.

\begin{figure}[!t]
\centering
\begin{tabular}{c c}
\includegraphics[width=0.45\linewidth]{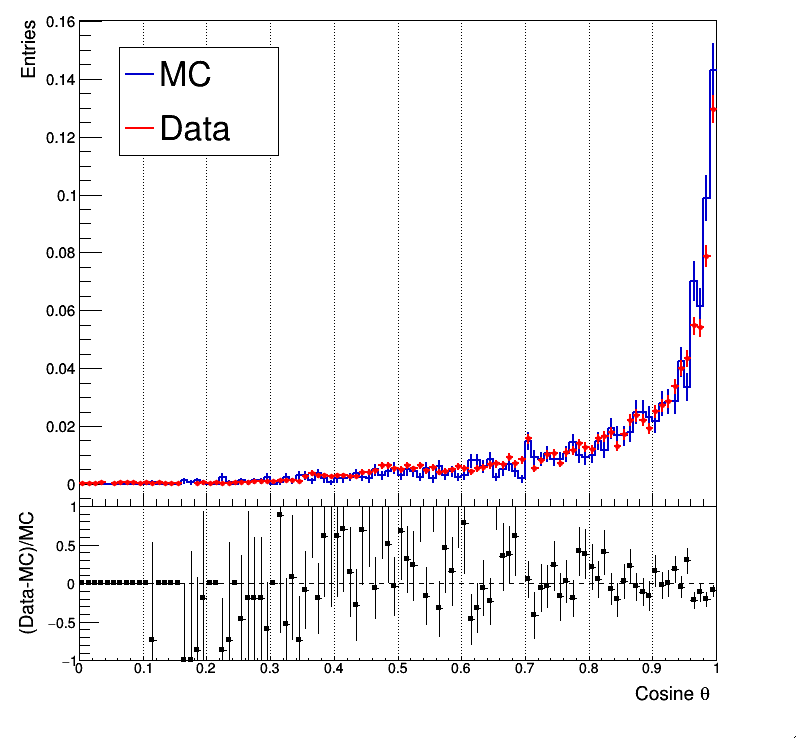} & \includegraphics[width=0.45\linewidth]{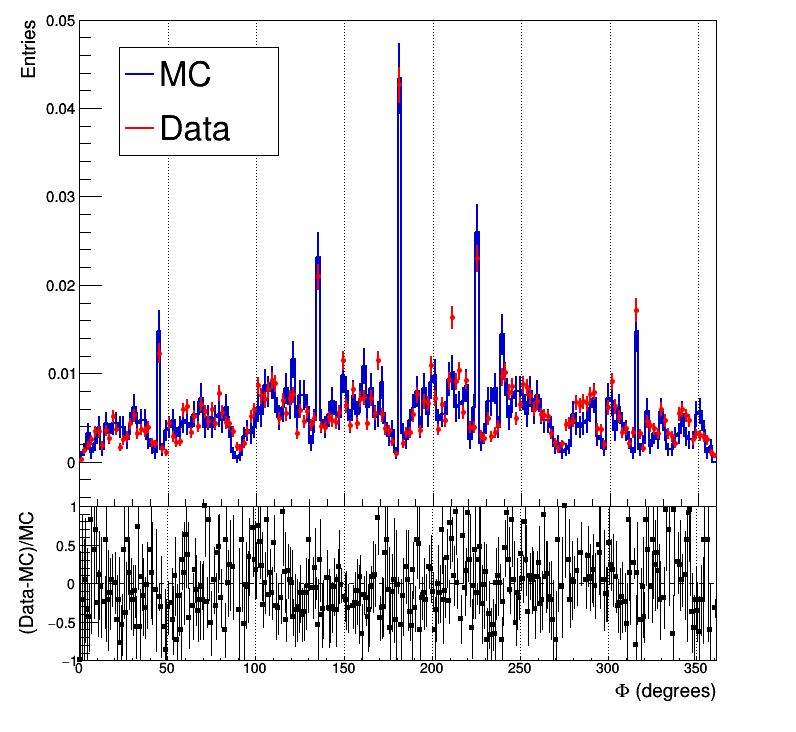}
\end{tabular}
\caption{Azimuthal (left) and polar (right) angle distribution of muon tracks in the SoLid prototype detector. Monte Carlo simulations are shown in blue, data in red.}
\label{fig:azimuthal}
\end{figure}

In Fig.~\ref{fig:azimuthal} the data versus Monte Carlo comparisons of both the azimuthal and polar angle distributions of reconstructed muon tracks are shown. The spikes in the polar angle distribution are introduced by the detector segmentation and muon track reconstruction in the analysis framework. These results show that muon reconstruction is well understood and that the simulation framework is trustworthy.



\subsection{Michel electrons and spallation neutrons}
\label{sec:spalneutron}

Cosmic muons can produce several secondary events in the detector. Two classes of those are treated in this section: spallation neutrons and Michel electrons. A correct reconstruction of these events validates the time synchronisation of the detector readout as well as the particle identification based on the pulse shape discriminant discussed earlier.

Michel electrons refer to muons that stop inside the detector and decay, $\mu^\pm \rightarrow e^\pm + \bar\nu_\mu (\nu_\mu) +  \nu_e (\bar\nu_e)$, producing a high energy electron or positron. 
Due to the minimal overburden of the prototype, a large amount of Michel electrons is expected. 

\begin{figure}[!t]
\centering
\begin{tabular}{l l}
\includegraphics[width=0.45\linewidth]{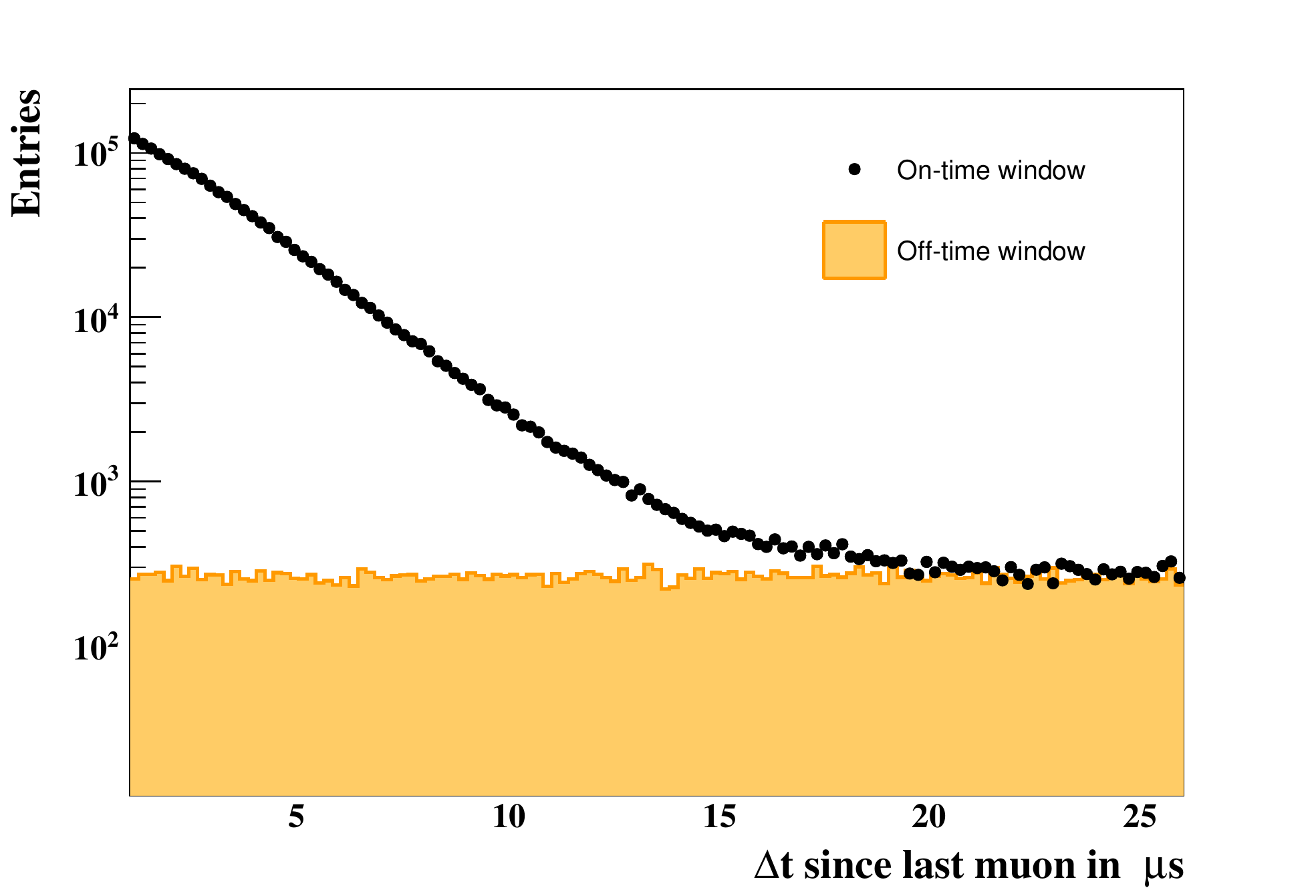} & \includegraphics[width=0.45\linewidth]{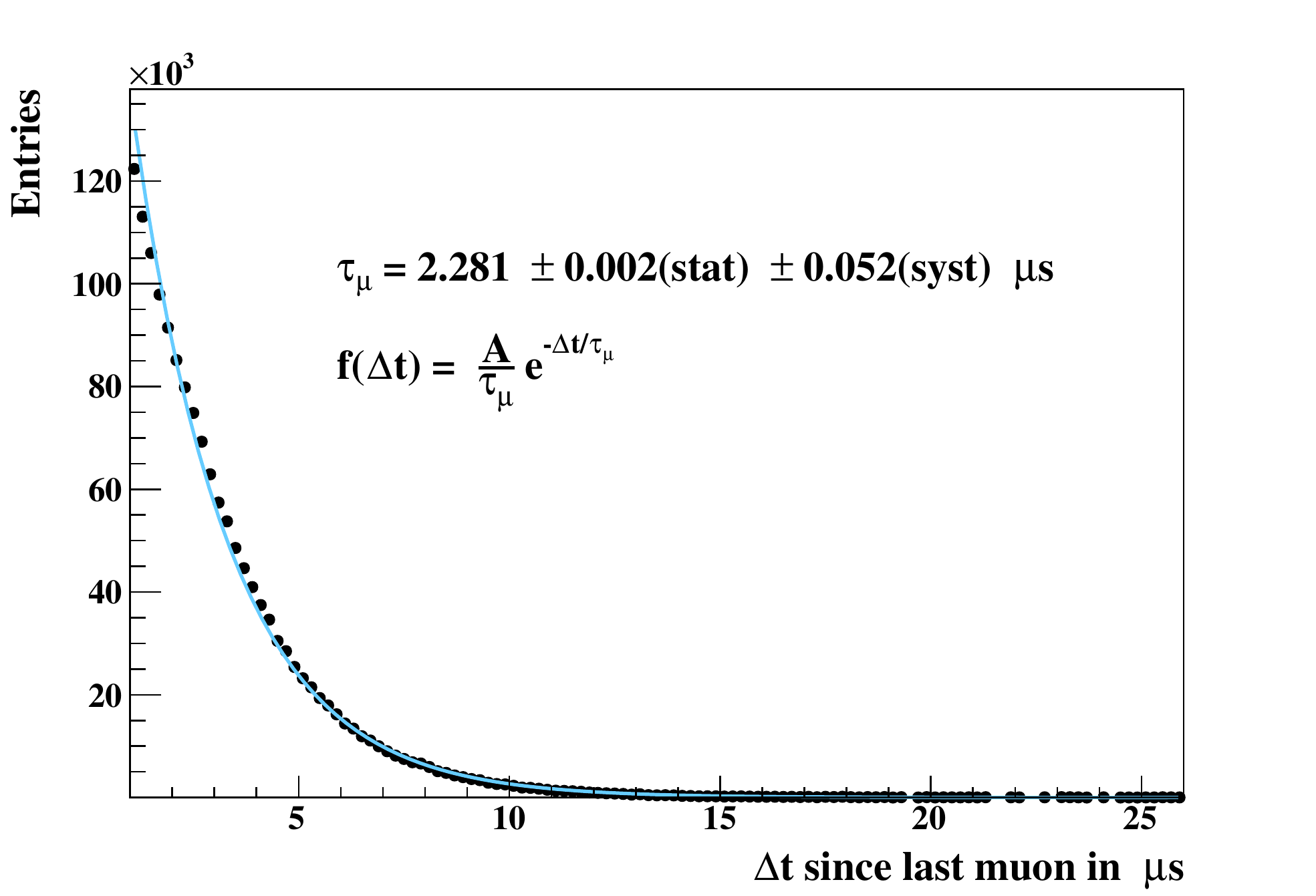}\\
\end{tabular}
\caption{$\Delta t$ distributions between a prompt muon and a Michel electron candidate for the on- and off-time windows (left). 
The on- minus off-time window distribution (right, background subtracted) was fitted with an exponential to get the muon lifetime ($\tau_\mu$). }
\label{fig:dts}
\end{figure}

The same classification of muons, EM and neutron-like signals as described in sections~\ref{sec:MuonID} and~\ref{sec:NeutronId} are used in what follows.
Stopped muons were selected searching for EM events with $E_{vis}> \,$3.5 MeV in a time window of 1 - 26~$\mathrm{\mu}$s after a tagged muon (on-time window). Owing to the characteristic decay time of muons this window appears to be suitable. Additionally, a  shifted window of 1.001 - 1.026~ms was also used to estimate the expected background (off-time window). Figure~\ref{fig:dts} (left) shows the time-difference between a prompt muon and a Michel electron candidate ($\Delta t_{\mu - e}$) for both on- and off-time windows. The $\Delta t_{\mu - e}$ distribution in the off-time window is flat, as expected.  
Subtracting the two data sets and fitting the remaining points with an exponential function, one finds the muon life-time value of $\tau_\mu = (2.281 \pm 0.002(stat) \pm 0.052(syst)) \, \mu s $, Fig.~\ref{fig:dts} (right). The systematic error was deduced by repeating the fit, excluding the first three points, and attributing the difference in $\tau_\mu$ as a 1$\sigma$ error.

\begin{figure}[!t]
\centering
\begin{tabular}{l l}
\includegraphics[width=0.45\linewidth]{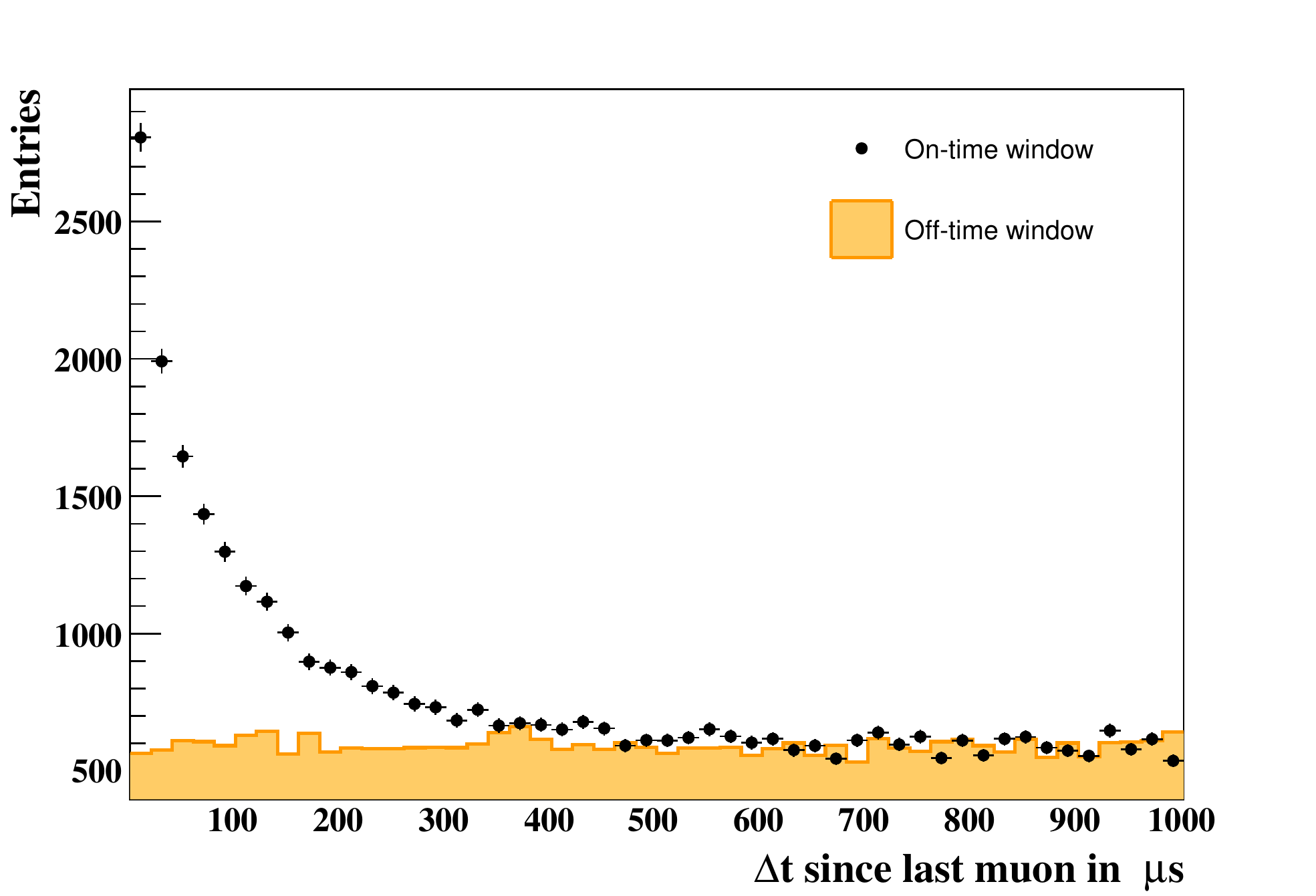} & \includegraphics[width=0.45\linewidth]{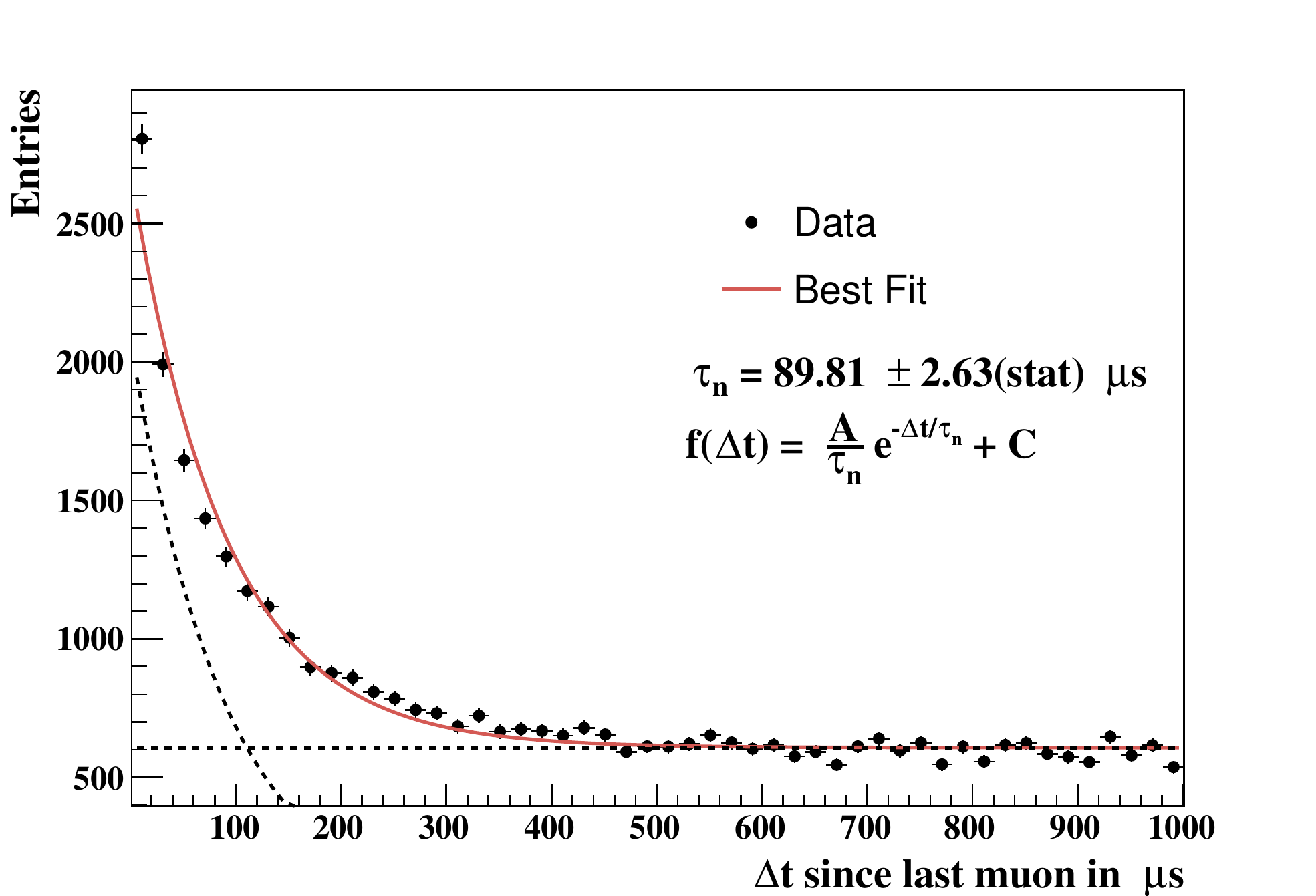}
\end{tabular}
\caption{$\Delta t$ distributions between a muon and neutron-like signal (left). 
The neutron capture time in LiF:ZnS(Ag) was deduced using a simple model of an exponential with a flat background (right).}
\label{fig:nudts}
\end{figure}

On the other hand, spallation neutrons were identified by searching for identified neutrons in a time window of 1 - 1001$\,\mu s$ after a tagged muon. 
Again, a time window shifted by 1$\,$ms was employed for background estimation. 
The on- and off-time $\Delta t_{\mu - n}$ distributions are shown in Fig.~\ref{fig:nudts} (left). 
A combined fit with an exponential function and a flat background (Fig.~\ref{fig:nudts}, right) yields the value of $\tau_n = 89.81 \pm 2.63(stat) \, \mu s $ for the neutron capture time on $^6$LiF:ZnS(Ag). 
This value, the uncertainty of which is statistically limited,  is compatible with the measured capture time using  AmBe data described in section~\ref{sec:neutroncaptime}.

Both measurements, combined with our knowledge on the decay properties of muons and the thermalisation of neutrons in PVT, confirm the purity of the particle identification and the validity of the timing of the detector electronics and readout. It gives confidence to investigate the IBD like signatures in the detector, characterized by a prompt EM signal, followed by a neutron induced signal in the $^6$LiF:ZnS(Ag) in a time window in scale with the measured neutron capture time.

\begin{figure}[!t]
	\centering
	\includegraphics[width=0.9\linewidth]{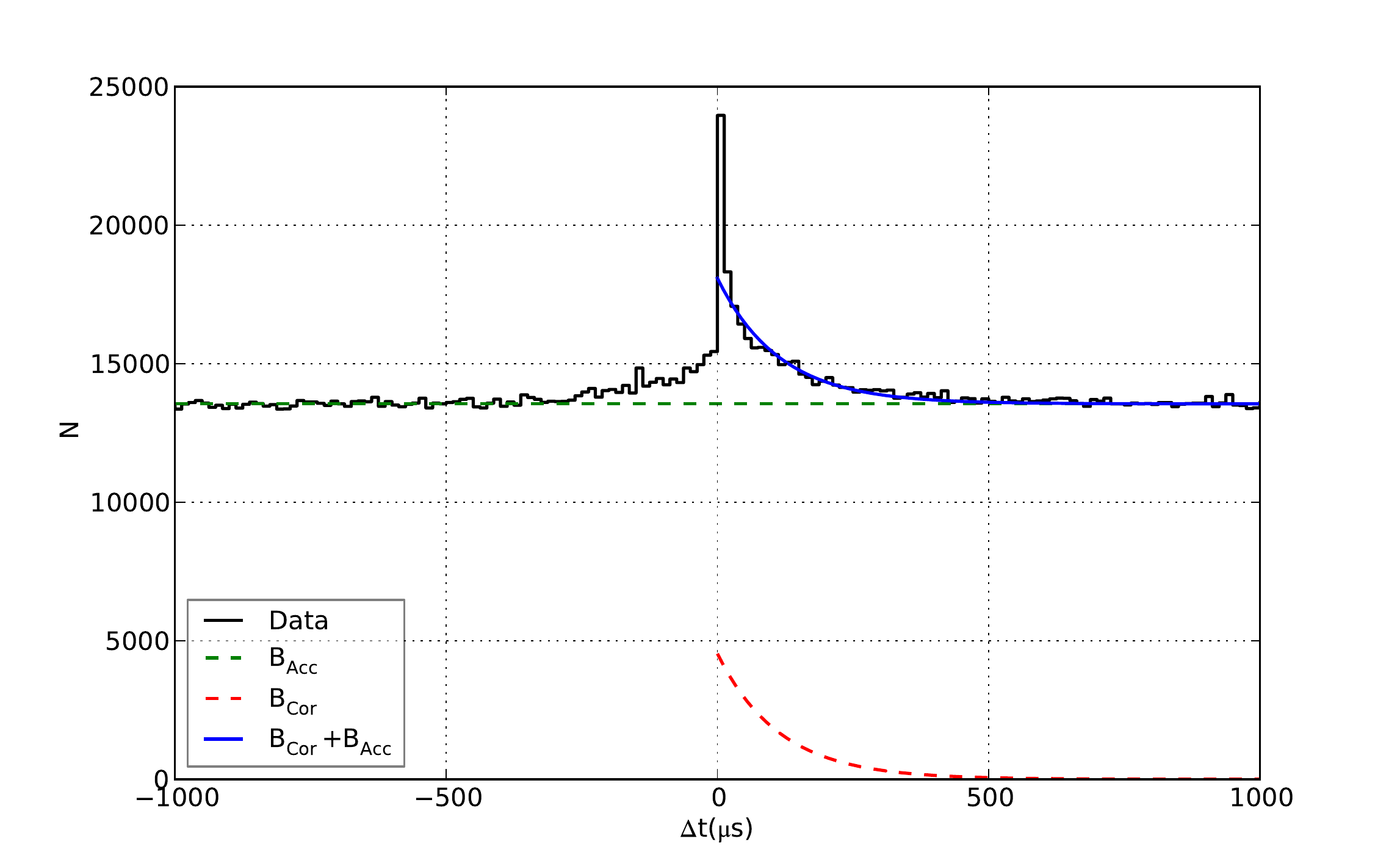}
	\caption{Distribution of the time difference between prompt and delayed events for IBD candidates, prior to any IBD selections for reactor-off data. An exponential fit to the correlated component is plotted for visual purposes.}
	\label{fig:deltrecon}
\end{figure}
\section{IBD analysis}
\label{sec:coincidence}
IBD candidates are found by searching for time correlations between prompt EM and delayed neutron-like signals in the detector, using the same identification criteria as discussed before. The time difference between these coincidence pairs, prior to the application of further selection criteria is shown in Fig.~\ref{fig:deltrecon} for a reactor off data sample.
It will allow to characterize the time-correlated and uncorrelated background components from data. We will assume, in a simple background model, that the time-correlated background rate and its visible energy spectrum remains unchanged during reactor operations. This is justified when no fast neutron signals are induced by the reactor operating at full power. The accidental background rate and energy spectrum will receive an additional component during reactor operations, but we will demonstrate here that it can be reduced to a small contribution by exploiting the spatial segmentation of the detector. Note that due to its flat time structure the uncorrelated background can always be determined by an off-time search window in reactor on and off conditions.
\begin{figure}[!t]
	\centering
	\includegraphics[width=0.5\linewidth]{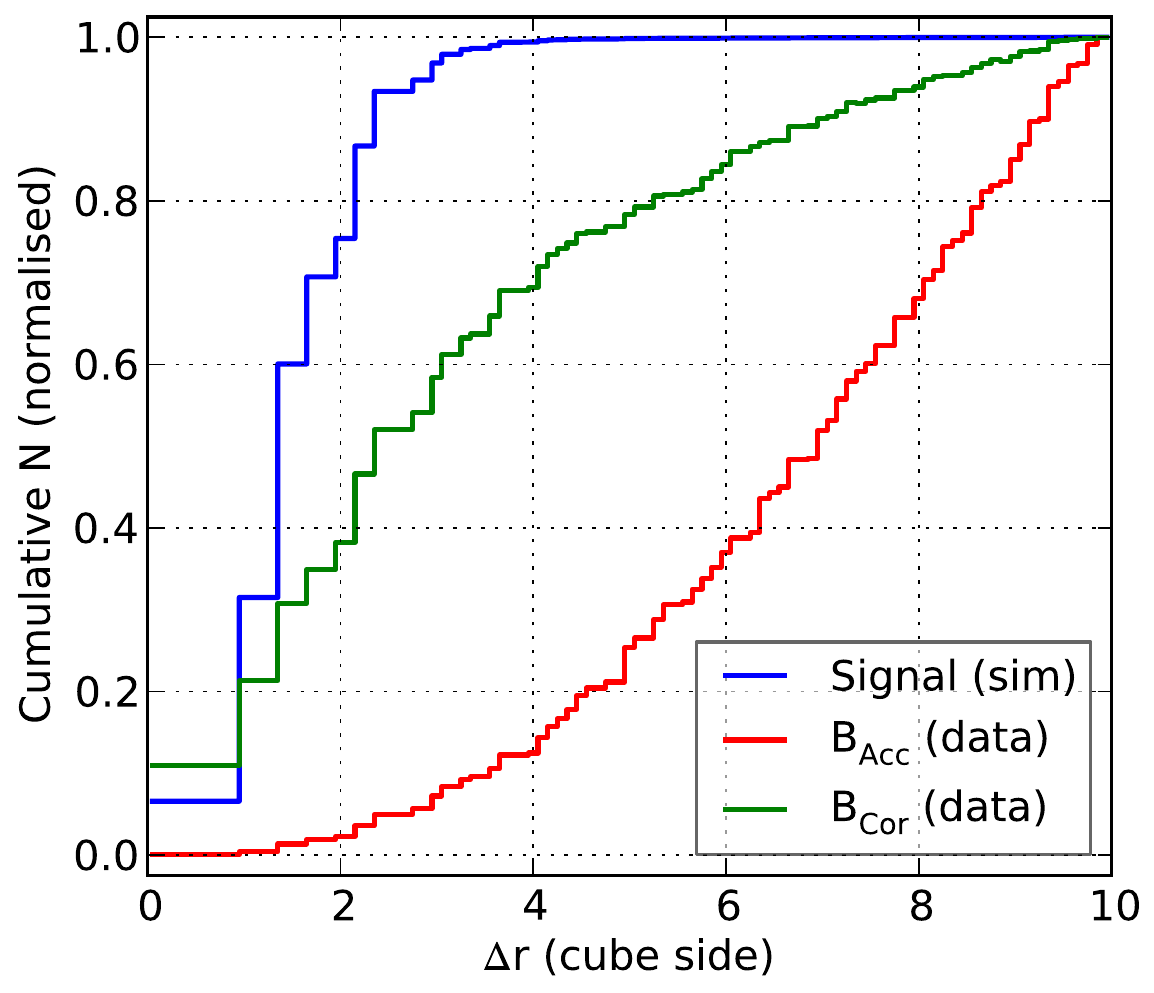}
	\caption{Cumulative distribution of $\Delta r$ (the radial separation between the positron and neutron candidates), prior to other IBD selections. The accidental background is for reactor on data.}
	\label{fig:delr}
\end{figure}

From Fig.~\ref{fig:deltrecon}, we deduce three main features, the rate of events being arbitrary:
\begin{itemize}
	\item a flat component is visible around N = 14000, for both positive and negative $\Delta t$, corresponding to the accidental background, $B_{acc}$;
	\item a large correlated component at positive $\Delta t$, which is referred to as the time-correlated background, $B_{Cor}$;
	\item a smaller correlated component at negative $\Delta t$, which is not contributing to the final positive time window used for the IBD search. 
\end{itemize}
\begin{figure}[!t]
	\centering
	\includegraphics[width=0.7\linewidth]{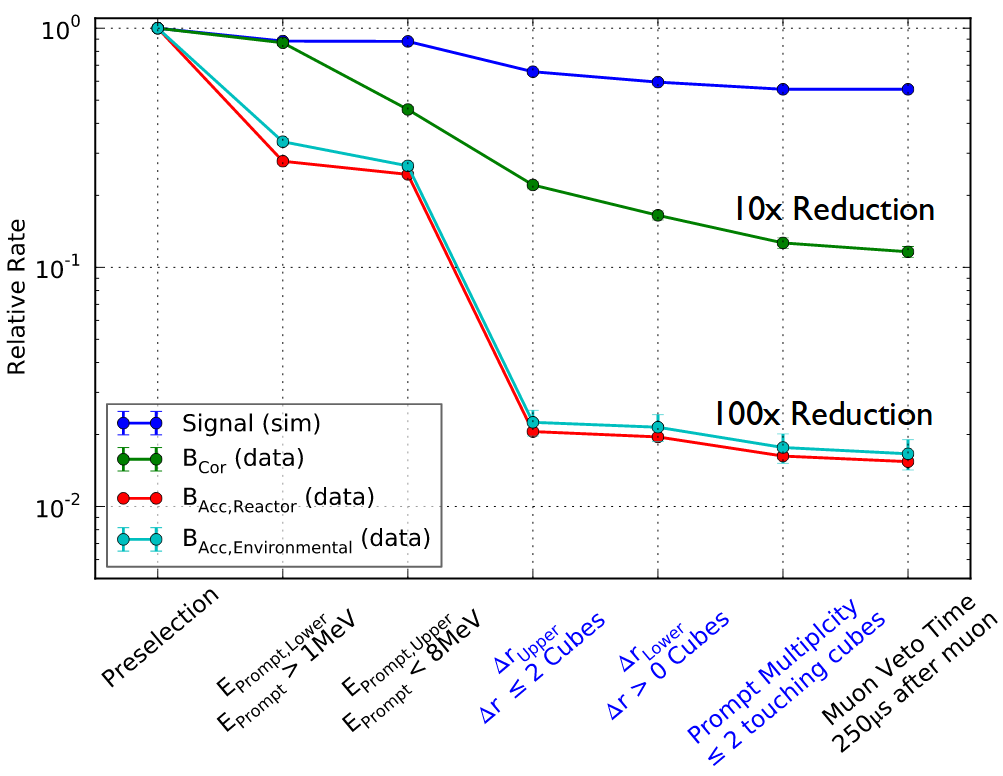}
	\caption{Signal and background relative rates for each selection cut applied sequentially. The order has been chosen such that selections on topological variables are applied last (highlighted in blue). The relative rates are obtained by normalising  to the number of prompt-delayed coincidences reconstructed (i.e. using a $\Delta t$ cut only).}
	\label{fig:cutsSequential}
\end{figure}
The last effect has been observed by other experiments~\cite{Bowden}, and corresponds to cosmic neutron showers. These showers can involve multiple fast neutrons interacting simultaneously with the detector. This contribution is interpreted as cases where neutrons are incorrectly correlated with prompts produced by other neutrons from the same shower. We will illustrate how the rate of the time-correlated and accidental backgrounds can be drastically reduced with further selection criteria, some of which are unique for the SoLid detector due to its high level of segmentation.

Based on the earlier measurements of the neutron thermalisation and capture time in sections~\ref{sec:neutroncaptime} and ~\ref{sec:spalneutron}, IBD signatures are formed for cases where the time difference between an EM signal and tagged neutron is smaller than $\Delta t < 220\,\mu$s. It contains $91\,\%$ of all time-correlated background events. For each of these retained coincidence pairs, the radial separation in terms of cubes, $(\Delta r)^2 = (\Delta xy)^2 + (\Delta z)^2$, is determined. The cumulative distribution of $\Delta r$ is shown in Fig.~\ref{fig:delr} for a simulated IBD signal and background components determined from the reactor-off data sample. It can be seen that the background populations extend to higher values of $\Delta r$, particularly for the accidental background, whereas the signal is mostly contained within a small number of cubes. We require further that the prompt EM and the delayed neutron-like signal are separated by at least one and maximally two cubes, which vastly reduces the accidental background component. 

Since the prompt energy of background coincidence pairs is shifted towards low energies, the IBD candidates are required to have a visible energy associated with the prompt EM signal in the range of $1<E_{Prompt}<8$$\,$MeV, where the upper bound serves as a muon veto. A multiplicity selection is also applied: the prompt candidate must be localised to maximally two cubes which share a cube face. This selection is effective at reducing muons or multiple proton recoils. Finally, a muon veto is applied, where IBD candidates cannot be formed within a time window of 250$\,\mu$s after a muon candidate. A summary of the effect of these IBD selection criteria, applied in succession, for reducing the background components is shown in Fig.~\ref{fig:cutsSequential}. Relative to forming coincidence pairs using timing information alone, referred to as pre-selection level, the accidental background has been reduced by more than a factor of 50, and the correlated background reduced by a factor of 10. The relative signal efficiency found for this selection from simulation, is $57\,\%$, and is specific for this prototype module. The accidental background under reactor-on conditions, determined by a shifted off-time window, is shown to exhibit the same behaviour as the environmental accidental background.
\section{Reactor data}
\label{sec:reactordata}
\begin{figure}[!t]
	\centering
	\includegraphics[width=0.7\linewidth]{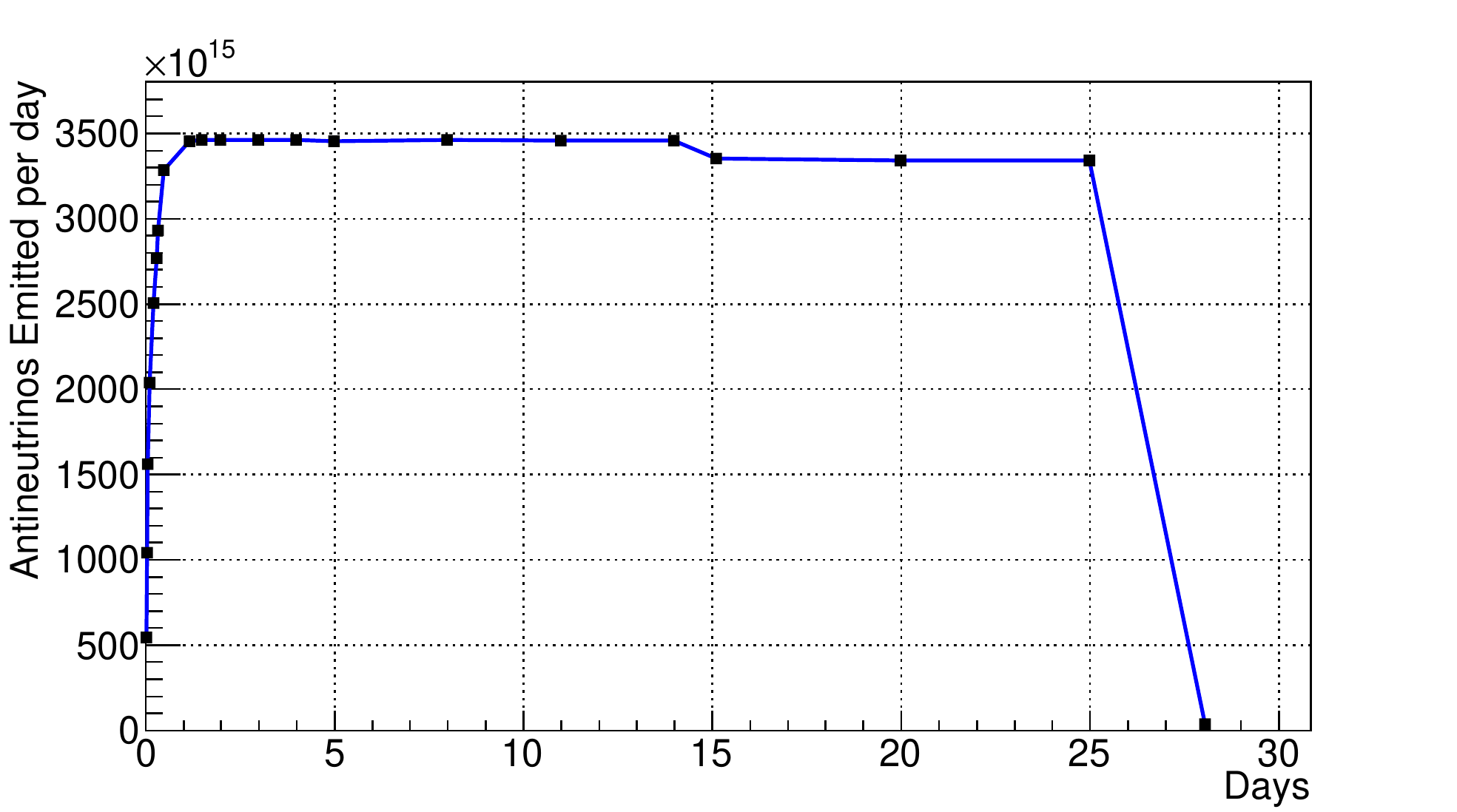}
	\caption{Emitted anti-neutrino flux calculated with the conversion method.}
	\label{fig:reactor}
\end{figure}
As mentioned earlier, the BR2 reactor overhaul and maintenance schedule coincided with the deployment of our submodule prototype. This allowed for only a small dataset to be collected of 50 hours under stable running conditions of the detector during reactor operations, as indicated in Tab.~\ref{tab:datasets}. Given the small neutron trigger and detection efficiency reported in section~\ref{sec:neutroneff}, the expected number of observed anti-neutrino events in the data run is $10 \pm 1 (stat)$ which is unlikely to result in any statistically significant anti-neutrino excess in the data. The data can nevertheless be used to validate the assumptions made in the simple background model described above. Compared to the current estimation of the accidental backgrounds, time-correlated backgrounds are the biggest concern for the success of the full-scale experiment. The different components of the time-correlated background will be discussed in the last section of this paper.

\subsection{Predicted anti-neutrino rate}
The accurate prediction of the flux and energy spectrum of electron anti-neutrinos is an involved process, subject to intense debates~\cite{PhysRevD.83.073006,Mueller:2011nm,Huber:2011wv,Hayes:2015yka, Huber:2016fkt, Huber:2016xis, Dentler:2017tkw}. The strategy of the SoLid collaboration is to use state-of-the-art methods to predict the anti neutrino flux and spectra, based on a detailed 3D model of the BR2 core coupled to MCNPX/CINDER'90~\cite{Silvapaper} as a starting point to produce the fission rates. In addition, the MURE code will be used to track the burn up of the fissile products in the core as prescribed by~\cite{five}. The SoLid experiment will have the option of computing a conversion spectrum based on Huber~\cite{Huber:2011wv} and Mueller~\cite{PhysRevC.83.054615}, supplemented by off equilibrium corrections provided by MURE~\cite{five}, or to use predictions using the summation method developed by the Nantes group~\cite{PhysRevLett.109.202504, Bui:2016otf}. 

The anti-neutrino rates for the BR2 reactor cycle covered in this paper are calculated using the conversion method based on the fission rates obtained with MCNPX/CINDER'90 convolved with the Huber's anti-neutrino converted spectra \cite{Huber:2011wv}. A preliminary calculation of the flux done with the converted method is presented in Fig.~\ref{fig:reactor} for this cycle. The time steps for the reactor evolution were chosen to take into account the power variation but also the needs of the reactor calculation.

The simulated anti-neutrino flux originating from the BR2 core is subsequently transmitted to the detector simulation code to obtain the real detected anti-neutrino rate. The geometrical acceptance of the prototype submodule amounts to 0.17\% for a surface exposure of 6400 cm$^2$. Convoluted with the neutron detection efficiency and the selection criteria described above, except for the  $\Delta r$ requirement, results in a predicted signal yield of 10$\pm$1 events. The rest of the background expectations and observed number of events during the 50 hours of reactor-on data are given in Tab.~\ref{tab_rate_cuts_on}.

\begin{table}[!b]
	\begin{center}
		\begin{tabular}{cc}
			\hline
			\hline
			Component & N IBD Candidates (scaled to Reactor On live time) \\
			\hline
			$B_{Off}$ & 258 $\pm$ 5 \\
			$B_{Acc,\,Environmental}$ & 25 $\pm$ 1 \\
			$B_{Cor}$ & 234 $\pm$ 5 \\
			\hline 
			$B_{Acc,\,On}$ &  61 $\pm$ 2 \\
			$B_{Acc,\,Reactor}$ & 37 $\pm$ 2 \\
			\hline
			$\langle S \rangle$ & 10 $\pm$ 1 \\  
			$\langle B_{On} \rangle$ & 295 $\pm$ 5 \\
			$\langle S:N \rangle$ & 1:30\\
			\hline
			$N_{Reactor\,On}$ & 305 \\
			Excess & 10 $\pm$ 18 \\
			Significance & 0.6$\sigma$ \\
			\hline
			\hline  
		\end{tabular}
		\caption{Expected Background yields, determined from reactor on and off periods, and expected signal yield based on reactor flux calculations and detector simulations, together with the observed excess and its statistical significance. All uncertainties are statistical only.}
		\label{tab_rate_cuts_on}
	\end{center}
\end{table}

\begin{figure}[!t]
	\centering
	\begin{tabular}{l l}
		\includegraphics[width=0.45\linewidth]{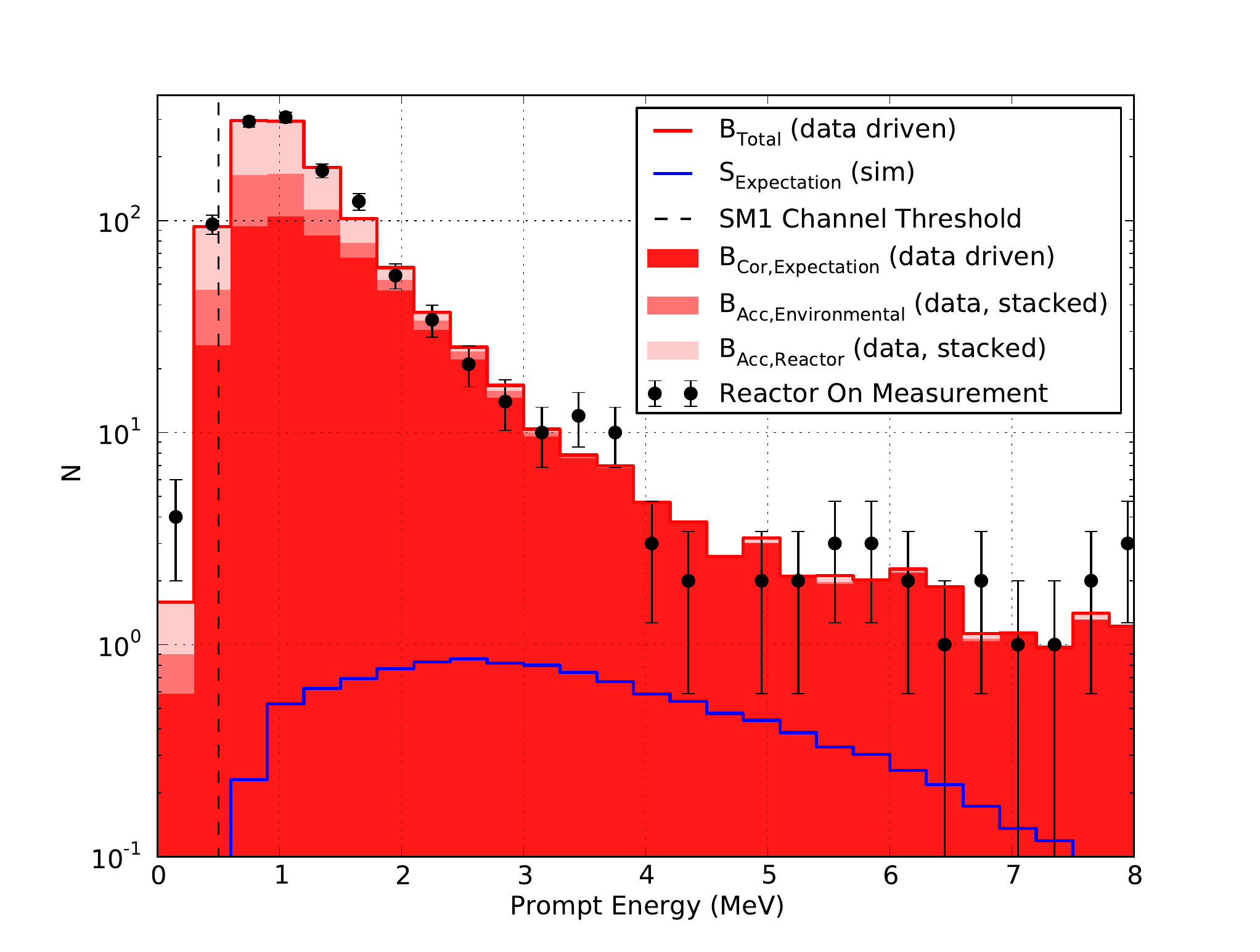} & \includegraphics[width=0.45\linewidth]{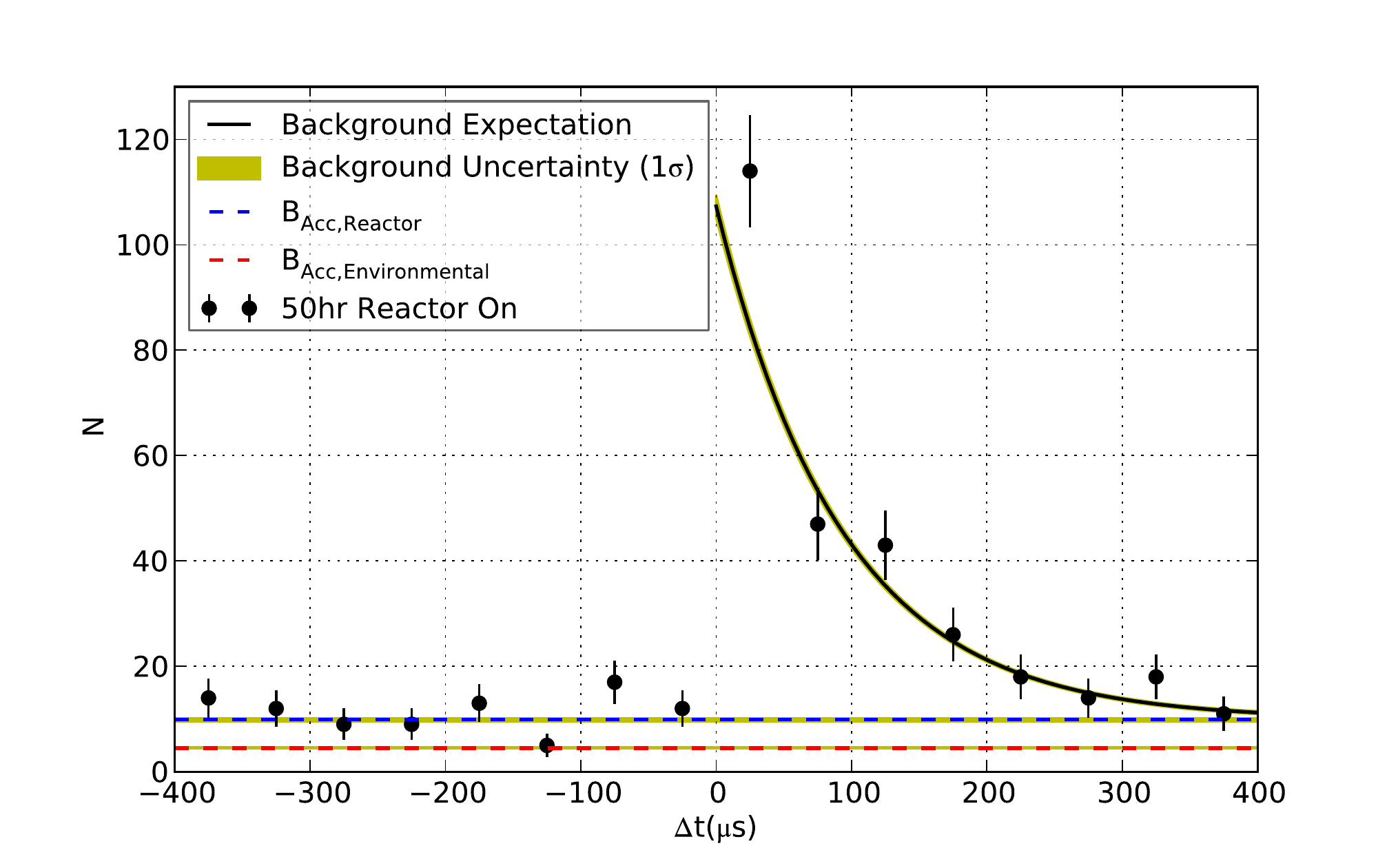}\\
	\end{tabular}
	\caption{Comparison between data and background model for the prompt energy distribution of selected IBD candidates (left), and for the $\Delta t$ distribution (right). }
	\label{fig:enPromptOn}
\end{figure}

Figure~\ref{fig:enPromptOn} shows a comparison between the reactor-on data and background model for the prompt visible energy distribution and the time difference, $\Delta t$, between the prompt signal and the delayed neutron candidate. The small signal expectation, from simulation, is also shown in the energy spectrum. The data agree with the background prediction, suggesting the background model is appropriate. 
Comparing the reactor on run to the expectation of the background, a small excess is observed. Coincidentally, the excess is very near the expected signal yield, but given the errors, this is not statistically significant.

\section{Time-correlated background analysis}
\label{sec:corrbg}
Most of the correlated background can be determined using the data collected during the reactor off period. Examples are the atmospheric neutrons in cosmic showers, spallation neutrons created by muons, and natural radioactivity from the decay chains of $^{238}$U and $^{232}$Th contaminating the detector material. In particular the neutron detection screens containing $^6$Li isotopes can be contaminated with trace fractions of $^{214}$Bi from the $^{238}$U decay chain. The $^{214}$Bi decays to $^{214}$Po while emitting a $\beta$ or $\gamma$'s with an energy up to the end-point of $Q_{\beta} = 3.27$$\,$MeV. $^{214}$Po has a half-life of $T_{1/2} = 163.6 \pm 0.6\,\mu$s, similar to the time for neutron thermalisation and capture, and emits an $\alpha$-particle with an energy of 7.69$\,$MeV when decaying to $^{210}$Pb~\cite{NuclearDataSheets2014}. The $\beta$ or $\gamma$ from the $^{214}$Bi decay gives rise to a prompt EM signal, while the $\alpha$ from the subsequent $^{214}$Po decay excites the ZnS.

\subsection{Observation of $^{214}$Bi decay}
\label{sec:BiPo}
In order to measure the $^{214}$Bi induced background, further denoted as BiPo, dedicated selection requirements are applied to the data collected when the reactor was off. The following selection requirements are chosen to select a pure sample of BiPo events: 
\begin{itemize}
\item Events with a muon preceding the ZnS signal are vetoed when $\Delta t = t(n-like) - t(\mu) <1.5$~ms. 
\item The ZnS and PVT signal are required to be observed in the same cube, $\Delta r(EM,n-like)=0$. 
\item The time between the ZnS and PVT signal is required to be less than 1~ms, $\Delta t = t(n-like) - t(EM)< 1$~ms. 
\item The energy of the PVT signal is required to be $0.6<E_{prompt}<3$~MeV. The upper limit is justified by the end-point energy $Q_{\beta}$.
\end{itemize}

\begin{figure}[!t]
	\centering
	\begin{tabular}{l l}
		\includegraphics[width=0.45\linewidth]{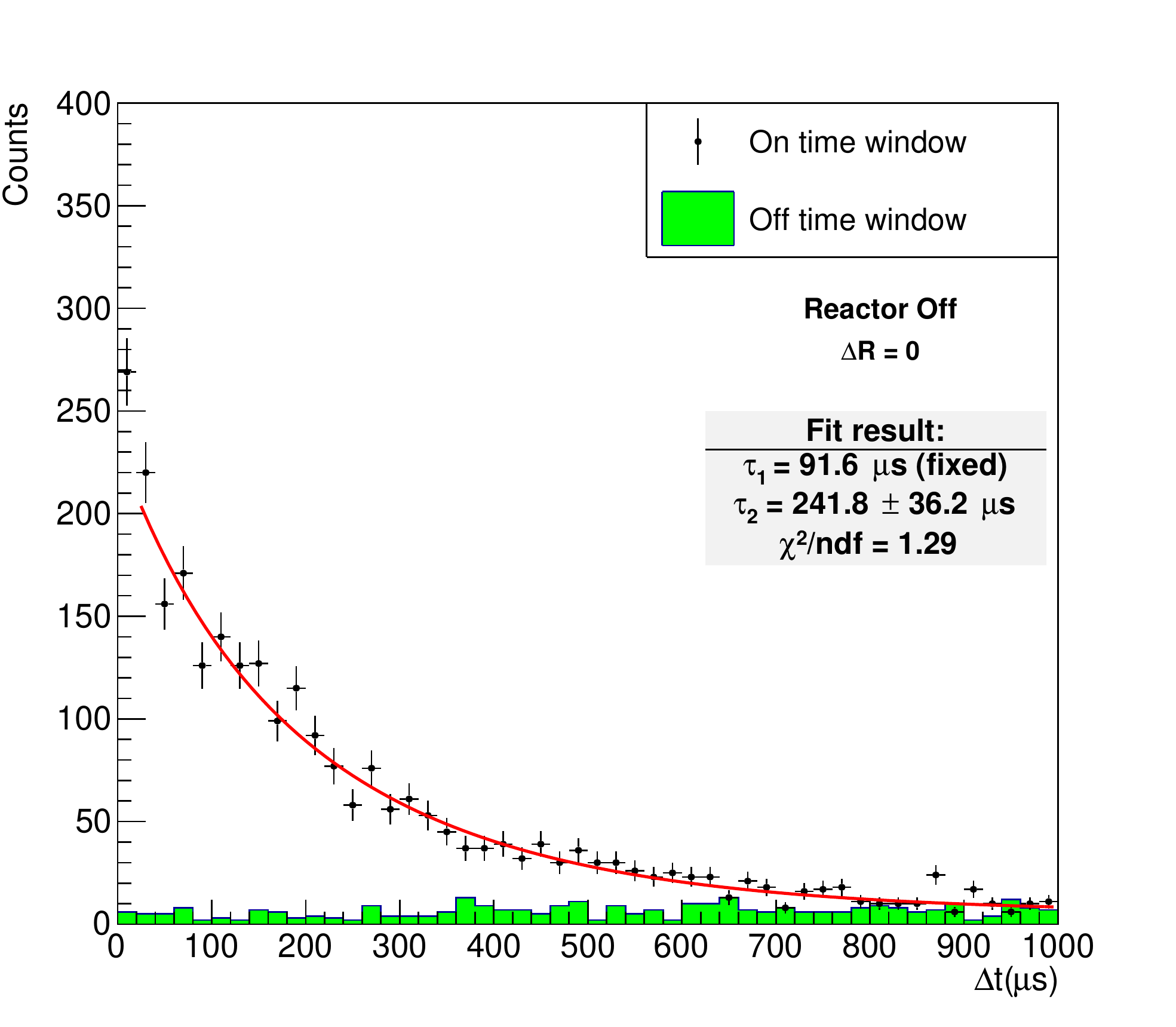} & \includegraphics[width=0.45\linewidth]{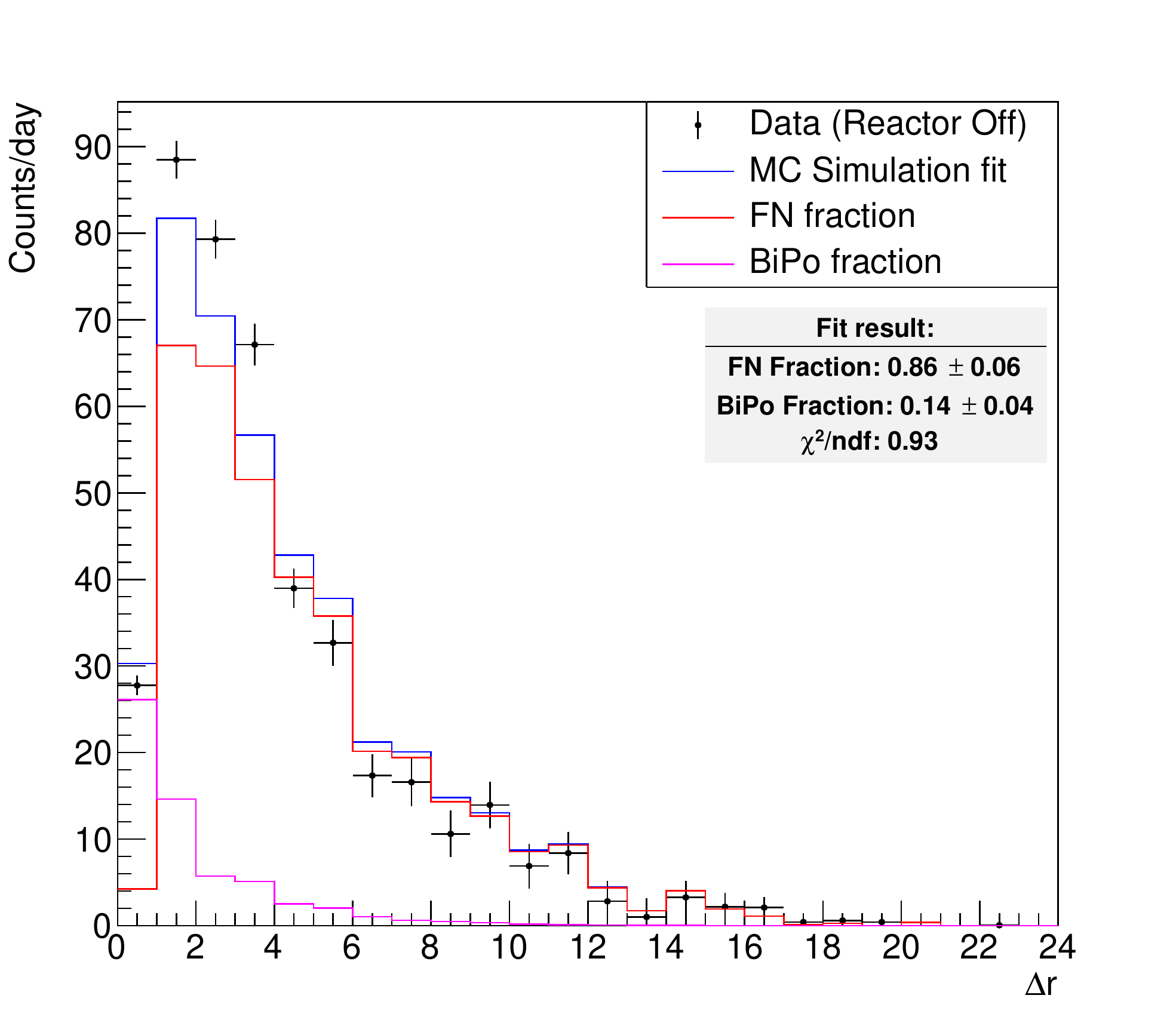}\\
	\end{tabular}
	\caption{Left: The distribution of the time difference between the ZnS and the PVT signal, $\Delta t$. A fit is applied to determine the fraction of BiPo events. Right: The distribution of the distance between the ZnS and PVT signal, $\Delta r$. A fit to the data is performed using the simulated shapes for the combined atmospheric and spallation neutron background (FN) and the BiPo background.}
	\label{fig:FitDeltat}
\end{figure}

The distribution of $\Delta t$ is shown in the left panel of Fig.~\ref{fig:FitDeltat}. 

The distribution is fitted in the range between $20$ and 1000$\,\mu$s with the analytical function:
\begin{equation}
f(\Delta t) = c_1 \times {\mathrm{exp}}(-\frac{\Delta t}{\tau_1}) + c_2 \times {\mathrm{exp}}(-\frac{\Delta t}{\tau_2})+c_{\mathrm{acc}}
\end{equation}
The decay time of the first exponential, $\tau_1$, is fixed to the value of the fast neutron thermalisation and capture time discussed in section~\ref{sec:neutroncaptime}. The fitted value of the constant $c_{\mathrm{acc}}$ corresponds to the accidental background component, while the decay time of the second exponential is directly related to the half-life of $^{214}$Po and measured to be $\tau_2=(241.8 \pm 49.7)~\mu$s. The uncertainty includes the systematic uncertainty that is estimated as the difference in the fitted $\tau_2$ value when fitting between 0  and 1000$~\mu$s. The effect of varying $\tau_1$ by its uncertainty is negligible compared to the other sources. The value of $\tau_2$ corresponds to a half-life of $T_{1/2} = \tau_2{\mathrm{ln}}2 = (167.6 \pm 34.4)~\mu$s, consistent with the value expected for $^{214}$Po. After the accidental subtraction, the fitted fraction of events from cosmic or spallation neutrons, $(8.2 \pm 10.9)\,\%$ is about a factor 10 smaller than for BiPo events and negligible in this study. This indicates that the sample is indeed very pure in BiPo events.

\subsection{Breakdown of the correlated background}
\label{sec:corrback}

To estimate the correlated background for the IBD event selection, the IBD selection requirements are applied to the reactor-off data, with the exception of the topological cut on the distance between the prompt EM and the delayed neutron-like signal, $\Delta r$. The accidental background is estimated from the data collected when the reactor is off, using a time window of the same size, but shifted by $-450$$\,$$\mu$s. The fraction of each correlated background process is then estimated by fitting the simulated $\Delta r$ distribution for the $^{214}$Bi$\,\rightarrow\,$$^{214}$Po as well as for the combined atmospheric and spallation neutron background to the data, after subtracting the accidental background contribution. This fit is shown in the right panel of Fig.~\ref{fig:FitDeltat}.
The fitted fractional rates of BiPo and atmospheric and spallation neutron events with $1\leq\Delta r<2$ provide a good estimate  of these backgrounds to the  observed event rate when the reactor is on. These results indicate that the time-correlated background for the IBD selection is dominated by atmospheric and spallation neutron events, contributing to about 82\% of the total number of observed events. For the BiPo events, a rate of $0.0073 \pm 0.0004 (stat.) \pm 0.0021 (syst.)$ counts/cube/day is obtained. 

The rate of BiPo events is also determined from the rate of these events passing the selection requirements in the previous section, yielding a very pure sample. Correcting for the different efficiencies corresponding to the two sets of event selection requirements, we obtain two rate predictions. The difference between this extrapolated rate and the rate obtained with the fit of the $\Delta r$ distribution is quoted as a systematic error on the BiPo event rate. 

The purification of the BiPo sample, as described in the previous section, can also be used as a veto criterium to reduce the contamination of intrinsic detector backgrounds in the full scale experiment, if needed.

\section{Conclusion and outlook}
The SoLid collaboration designed and constructed a full scale prototype module of a new reactor neutrino detector 
with a per-mille level of control on its proton content. The prototype was deployed and commissioned at the intended location of a large scale detector of 1.6 ton, where it demonstrated excellent operational stability over a period of several months.

The commissioning of the module demonstrates the equalisation and calibration of the individual channel energy response up to a level of 3\% and an energy resolution of 20\% for 1 MeV electrons interacting in the plastic scintillator, that can be improved to 14\% by means of several design improvements. Calibration runs taken with gamma and neutron sources demonstrate the possibility to exploit pulse shape discrimination for particle identification. The analysis of cosmic ray data demonstrates the excellent muon tracking capabilities of the detector which can be used to calibrate the energy response and monitor its stability over time.

The prototype collected useful data during reactor operations, albeit with suboptimal trigger performance due to unexpected features in the readout electronics. Event reconstruction and analysis tools were developed to perform physics analysis with the collected data. Using this data, accidental and time-correlated backgrounds are measured in realistic conditions. 

The spatial segmentation, a unique feature  of the SoLid detector, is shown to be a powerful tool in reducing both accidental and time-correlated backgrounds. The background model, which is completely data driven, shows its applicability to the IBD analysis. The components of the time-correlated background are analysed using dedicated selections and are extrapolated to the IBD search region, indicating that fast neutron induced backgrounds dominate the experimental uncertainties in the current set-up. These results are taken into account in the shielding strategy of the full-scale detector.

Based on the observations made with the prototype module, the SoLid collaboration adapted a set of design changes for the full scale experiment. These include a new design of the front-end electronics with far superior noise-tolerances, and an improved light collection to achieve a better energy resolution. In addition the final detector will contain twice the amount of $^6$LiF:ZnS(Ag) screens to increase the neutron detection efficiency. The detector will also be cooled to an ambient temperature of 5$^\circ$$\,$C to reduce the dark count rate, in combination with a passive shielding to reduce the cosmic and reactor induced backgrounds. Finally, a dedicated neutron trigger based on pulse shape discrimination will be deployed.

A year of construction results in a very improved experiment consisting of 12800 detection cells, corresponding to a fiducial mass of 1.6 ton which is currently being commissioned near the BR2 reactor.
The commissioning and performance of the full-scale experiment will be discussed in a following paper.

\clearpage
\section{Acknowledgements}

This work was supported by the following funding agencies: Agence Nationale de la Recherche grant ANR-$16­\mathrm{CE}31­0018­03$, Institut Carnot Mines, CNRS/IN2P3 et Region Pays de Loire, France; FWO-Vlaanderen and the Vlaamse Herculesstichting, Belgium; The U.K. groups acknowledge the support of the Science \& Technology Facilities Council (STFC), United Kingdom; We are grateful for the early support given by the sub-department of Particle Physics at Oxford and High Energy Physics at Imperial College London. We thank also our colleagues, the administrative and technical staffs of the SCK\raisebox{-0.8ex}{\scalebox{2.8}{$\cdot$}}CEN for their invaluable support for this project. Individuals have received support from the FWO-Vlaanderen and the Belgian Federal Science Policy Office (BelSpo) under the IUAP network programme; The STFC Rutherford Fellowship program and the European Research Council under the European Union's Horizon 2020 Programme (H2020-CoG)/ERC Grant Agreement \mbox{n. 682474} (corresponding author); Merton College Oxford.

\bibliographystyle{JHEP}
\bibliography{SoLid_paper2}

\end{document}